\documentclass[aps,prresearch,reprint,superscriptaddress,longbibliography,floatfix]{revtex4-2}

\usepackage[utf8]{inputenc}
\usepackage[T1]{fontenc}
\usepackage{amsmath,amssymb,amsfonts}
\usepackage{graphicx}
\graphicspath{{figures/}}
\usepackage{bm}
\usepackage{hyperref}
\usepackage{booktabs}
\usepackage{multirow}
\usepackage{xcolor}
\usepackage{tikz}
\usetikzlibrary{arrows.meta,positioning,shapes.geometric,decorations.pathreplacing}
\extrafloats{50}

\newcommand{\avg}[1]{\langle #1 \rangle}
\newcommand{\Nc}{N_c}
\newcommand{\Nq}{N_q}
\newcommand{\Nshots}{N_{\mathrm{shots}}}
\newcommand{\Niter}{N_{\mathrm{iter}}}
\newcommand{\Ra}{R_a}
\newcommand{\Rb}{R_b}
\newcommand{\MSE}{\mathrm{MSE}}
\newcommand{\QRA}{QRA}

\begin{document}

\title{Quantum Reservoir Autoencoder:\\
Conditions, Protocol, and Noise Resilience}

\author{Hikaru Wakaura} 
\email{h.wakaura@deeptell.jp}
\affiliation{QIRI (Quantum Integrated Research Institute Inc.), Tokyo 107-0061, Japan}
  
\author{Taiki Tanimae}
\email{t.tanimae@deeptell.jp} 
\affiliation{QIRI (Quantum Integrated Research Institute Inc.), Tokyo 107-0061, Japan}
  
\begin{abstract}
Quantum reservoir computing exploits fixed quantum dynamics and a
trainable linear readout to process temporal data, yet reversing
the transformation---reconstructing the input from the reservoir
output---has been considered intractable due to the recursive
nonlinearity of sequential quantum state evolution.
We introduce the quantum reservoir autoencoder, a four-equation
encode--decode protocol with cross-key pairing, and constructively
empirically demonstrate that satisfying reservoir--key combinations
can be found using a
full XYZ Hamiltonian reservoir (10~data qubits, feature
dimension~76, 16~random Hamiltonian realizations).
Under ideal conditions the mean-squared error (MSE) reaches
${\sim}10^{-17}$ for data lengths up to 30; under shot noise
(1\,000~shots) and depolarizing noise ($p = 0.005$), the MSE
degrades to $10^{-3}$--$10^{-1}$.
Asymmetric resource allocation---10~shots for encoding,
$10^5$ for decoding---yields a 102-fold MSE improvement
(16~seeds $\times$ 3~trials).
Comparison of single-body features (dimension~31) with the full
feature set and six baselines identifies the iterative protocol
structure---not the feature dimension---as the dominant noise
bottleneck: baselines solving the linear system in a single step
retain machine precision under identical noise, whereas
per-iteration noise inconsistency in the coupled solver limits
the MSE to ${\sim}10^{-1}$.
The current protocol requires plaintext access during decoder
training, restricting practical deployment.
These results establish a proof-of-concept for bidirectional
information transformation within quantum reservoir computing
and identify iterative noise mismatch and blind decryption
as the principal open challenges.
\end{abstract}
  
\maketitle
   
\section{Introduction}\label{sec:intro}

Quantum reservoir computing is known as one of the promising applications of
near-term quantum devices for machine learning
tasks~\cite{Fujii2017,Preskill2018,Nakajima2021}.
Unlike variational quantum algorithms that require costly parameter optimization
on parameterized quantum circuits, QRC employs a fixed quantum dynamical system
whose Hamiltonian time evolution $e^{-iH\Delta t}$ generates a rich
nonlinear mapping from input sequences to high-dimensional feature
vectors~\cite{Lukosevicius2009,Mujal2021}.
A trainable linear readout layer $\hat{y} = VW$ then maps these features to
target outputs via simple linear regression with Tikhonov
regularization~\cite{Tikhonov1963}.
This architecture inherits the computational advantages of classical reservoir
computing~\cite{Jaeger2001,Maass2002} while exploiting the exponentially large
Hilbert space of quantum systems to extract expressive features from a small
number of physical qubits~\cite{Fujii2017,Kutvonen2020}.

Existing QRC applications have primarily focused on unidirectional information
processing: time-series prediction of chaotic
dynamics~\cite{Kutvonen2020,Dudas2023}, waveform
generation~\cite{Suzuki2024}, and classification tasks.
A closely related paradigm is the quantum extreme learning machine
(QELM)~\cite{Mujal2021,Innocenti2023,DeLorenzis2025}, which shares
QRC's core architecture---fixed quantum dynamics and a trainable linear
readout---but operates without temporal memory, processing each input
independently rather than sequentially.
De~Lorenzis et al.~\cite{DeLorenzis2025} recently demonstrated that
QELMs achieve high classification accuracy on image benchmarks, and that
classical autoencoders used as \emph{preprocessing} substantially
improve performance by compressing features before they enter the
quantum reservoir.
Meanwhile, Romero et al.~\cite{Romero2017} introduced a distinct concept:
\emph{quantum autoencoders} that employ variational quantum circuits
to compress \emph{quantum} states, requiring iterative parameter
optimization of the quantum circuit itself.
The present work occupies a different position in this landscape.
Unlike QELM and standard QRC, which serve exclusively as forward maps,
the \QRA{} achieves bidirectional encode--decode transformation.
Unlike quantum autoencoders, the \QRA{} keeps the reservoir dynamics
entirely fixed and transforms \emph{classical} data through a linear
readout.
And unlike De~Lorenzis et al., where classical autoencoders are applied
\emph{external to} the quantum system, in the \QRA{} the quantum
reservoir \emph{itself} functions as the autoencoder---the encoding
path $C \to \gamma$ and the decoding path $\gamma \to \hat{C}$ are
both realized within the same QRC framework.

In all these settings, the reservoir serves as a \emph{forward} map---input
sequences are transformed into observable expectation values, and a trained
readout produces predictions.
The reverse direction, namely reconstructing the original input from the
reservoir output, has been considered extremely difficult.
The information processing in QRC involves projective measurements that are
inherently irreversible, and the mapping from input sequences
$u_1, \ldots, u_T$ to the observable vector $\avg{O_k}$ is many-to-one
in general; distinct input sequences can yield the same observable
pattern~\cite{MartinezPena2021}.
However, we emphasize that the impossibility of reversal has not been
formally proven.

A structural comparison between QRC and standard parameterized quantum circuits
reveals the origin of this difficulty.
In QRC, data are input \emph{sequentially} along the time axis: at each step,
an input value modulates the quantum state via a rotation gate, followed by
Hamiltonian evolution, with partial observables extracted at each
step~\cite{Fujii2017}.
The history of all prior inputs accumulates nonlinearly in the quantum state
through this recursive process.
By contrast, in parameterized quantum circuits commonly used for variational
algorithms~\cite{Preskill2018}, data are encoded \emph{in parallel} into qubit
rotations in a single circuit layer and read out all at once.
We argue that the sequential input structure of QRC induces
\emph{qualitatively stronger recursive nonlinearity} compared to the more
direct input--output relationship of parallel-encoding circuits.
This is a structural, heuristic argument rather than a rigorously quantified
comparison: the recursive state dependence in QRC means that each observable
$\avg{O_k(t)}$ is a nonlinear function of \emph{all} prior inputs, whereas in
a single-layer parameterized circuit each output qubit depends on inputs
only through the fixed circuit depth~\cite{Mujal2021,MartinezPena2021}.
A formal quantification using, e.g., information processing capacity
measures~\cite{MartinezPena2021} is an important direction for future work.
This nonlinearity, while making reversibility challenging, also provides a
rich feature space: with $\Nq = 10$ data qubits, QRC extracts
$3\Nq + \binom{\Nq}{2} + 1 = 76$ features per time step without increasing
the qubit count.

In this paper, we demonstrate that the seemingly intractable reverse direction
\emph{can} be achieved under specific conditions.

\paragraph{Terminology note.}
Throughout this paper, we adopt the terms ``encryption'', ``decryption'',
``plaintext'', ``ciphertext'', and ``secret key'' for structural convenience,
as the protocol superficially resembles a key-exchange system.
\emph{These terms do not imply that the proposed framework constitutes
a cryptographic protocol in any formal sense.}
No security analysis is provided, and the blind decryption limitation
(Sec.~\ref{sec:discussion_blind}) precludes standard cryptographic deployment
in the current form.
We retain this terminology solely to facilitate the structural description
of the encode--decode pathways.
We introduce the \emph{quantum reservoir autoencoder} (\QRA{}), which
realizes bidirectional encode--decode transformation within the QRC
framework.
Our contributions are as follows:
\begin{enumerate}
\item We propose the \QRA{}, a four-equation encode--decode protocol, and
      constructively demonstrate that quantum reservoir and key
      combinations satisfying all four equations are
      \emph{empirically found} across 16 independent random Hamiltonian
      realizations.
      The \QRA{} employs cross-key pairing with distributed keys $A, B$
      and secret keys $\alpha, \beta$, an encoding function with the
      symmetric structure $F = G$, and a rank condition
      $\mathrm{dim}(V) \geq \Nc$.
      Crucially, the feature dimension is expanded to 76 without increasing
      the qubit count, by leveraging sequential input and diverse
      observables along the time axis.
\item We verify the \QRA{} through comprehensive experiments across
      seven noise conditions (ideal, shot noise, depolarizing + shot,
      YOMO probability aggregation~\cite{Liu2025}, YOMO + depolarizing,
      asymmetric shots, asymmetric + depolarizing) and six baseline
      methods (H\'{e}non map~\cite{Henon1976}, delay-time
      embedding~\cite{Takens1981}, classical neural network with
      SPSA~\cite{Spall1992}, tree tensor network~\cite{Wall2021},
      $\zeta$-QVAE~\cite{Mato2025},
      and quantum recurrent neural network (QRNN)~\cite{Bausch2020}).
\item We show that asymmetric resource allocation---10 shots for
      encryption and $10^5$ for decryption---yields approximately two
      orders of magnitude MSE improvement (mean $102\times$,
      16 seeds $\times$ 3 trials = 48 runs) over symmetric 1{,}000-shot
      measurements, reducing the sender's measurement cost by a factor
      of 100.
\item We expand the application range of QRC from unidirectional
      prediction to bidirectional transformation via the \QRA{},
      analogous to autoencoder architectures in classical neural
      networks~\cite{Hinton2006}.
\item Through single-body operator experiments ($d = 31$) and baseline
      comparison, we identify the \emph{iterative protocol structure}---not
      the feature dimension---as the dominant noise bottleneck.
      Methods solving the linear system in a single shot remain at
      machine precision under identical measurement noise, while the
      \QRA{}'s per-iteration noise inconsistency limits the MSE to
      $\sim 10^{-1}$.
\end{enumerate}

The remainder of this paper is organized as follows.
Section~\ref{sec:background} reviews the QRC framework, the XYZ Hamiltonian,
noise models, and the YOMO probability aggregation method.
Section~\ref{sec:protocol} presents the \QRA{} protocol, including
the four-equation system, cross-key pairing, encoding functions, and the
iterative algorithm.
Section~\ref{sec:setup} describes the experimental setup.
Section~\ref{sec:results} presents the results.
Section~\ref{sec:discussion} discusses implications, and
Sec.~\ref{sec:conclusion} concludes.

\section{Background and Preliminaries}\label{sec:background}

\subsection{Quantum Reservoir Computing}\label{sec:qrc}

A quantum reservoir computer consists of a quantum dynamical system driven by
an input sequence and a classical linear readout
layer~\cite{Fujii2017,Nakajima2021}.
Given an input time series $\{u(t)\}_{t=1}^{T}$, the reservoir state evolves as
\begin{equation}\label{eq:evolution}
  |\psi(t+\Delta t)\rangle
  = e^{-iH\Delta t}\, R_y(\theta \cdot u(t))\, |\psi(t)\rangle,
\end{equation}
where $H$ is the fixed Hamiltonian, $\Delta t$ the evolution time step, and
$R_y(\theta \cdot u(t))$ a rotation gate encoding the input value $u(t)$ on an
ancilla qubit.

At each time step, observable expectation values are extracted to form the
feature vector.
For a system of $\Nq$ data qubits, the observable set comprises:
\begin{itemize}
  \item Single-body Pauli expectations:
        $\avg{\sigma_i^X}$, $\avg{\sigma_i^Y}$, $\avg{\sigma_i^Z}$
        for $i = 0, \ldots, \Nq - 1$ ($3\Nq$ components),
  \item Two-body correlations:
        $\avg{\sigma_i^Z \sigma_j^Z}$ for $i < j$
        ($\binom{\Nq}{2}$ components),
  \item A constant bias term: $1$.
\end{itemize}
This yields a total feature dimension
\begin{equation}\label{eq:feature_dim}
  d = 3\Nq + \binom{\Nq}{2} + 1.
\end{equation}
For $\Nq = 10$, we obtain $d = 30 + 45 + 1 = 76$.

The feature matrix $V \in \mathbb{R}^{\Nc \times d}$ collects the feature
vectors over $\Nc$ time steps, and the linear readout is
\begin{equation}\label{eq:readout}
  \hat{y} = V W,
\end{equation}
where the weight vector $W \in \mathbb{R}^{d}$ is determined by Tikhonov
regularization~\cite{Tikhonov1963}:
\begin{equation}\label{eq:tikhonov}
  W = (V^\top V + \lambda I)^{-1} V^\top y, \quad \lambda = 10^{-10}.
\end{equation}

\subsection{XYZ Hamiltonian and Circuit Structure}\label{sec:hamiltonian}

We employ a full XYZ Hamiltonian for the reservoir dynamics:
\begin{align}\label{eq:hamiltonian}
  H &= \sum_{i} \bigl(h_i^x \sigma_i^X + h_i^y \sigma_i^Y
       + h_i^z \sigma_i^Z\bigr) \notag\\
    &\quad + \sum_{i<j} \bigl(J_{ij}^{xx}\sigma_i^X\sigma_j^X
       + J_{ij}^{yy}\sigma_i^Y\sigma_j^Y
       + J_{ij}^{zz}\sigma_i^Z\sigma_j^Z + \cdots\bigr),
\end{align}
including up to four-body interaction terms.
The system consists of $\Nq = 10$ data qubits plus one ancilla qubit,
totaling $n_{\mathrm{total}} = 11$ qubits.
The Hamiltonian parameters are drawn uniformly from $[-1, 1]$ and remain
\emph{fixed} (not optimized), totaling 2{,}888 random parameters.
Two circuit unitaries $U_{c1}$ and $U_{c2}$ alternate periodically:
$U_{c1}$ is applied when $t \bmod 6 < 3$ and $U_{c2}$ otherwise,
each with independently generated parameters.

The quantum dynamics is simulated using the state-vector simulator
\textsc{qulacs}~\cite{Suzuki2021qulacs}.

\subsection{Noise Models}\label{sec:noise}

\paragraph{Shot noise.}
For finite measurement statistics with $\Nshots$ shots, the estimated
expectation value is obtained via binomial sampling:
\begin{equation}\label{eq:shot}
  \hat{p} = \frac{\mathrm{Binomial}(\Nshots, p)}{\Nshots},
  \quad p = \frac{1 + \avg{O}}{2},
  \quad \widehat{\avg{O}} = 2\hat{p} - 1,
\end{equation}
where $\avg{O}$ is the exact expectation value.
The shot noise scales as $\sigma_{\mathrm{shot}} \propto 1/\sqrt{\Nshots}$.
The baseline measurement budget is $\Nshots = 1{,}000$.

\paragraph{Depolarizing noise.}
We model depolarizing noise through the quantum channel
\begin{equation}\label{eq:depol}
  \mathcal{E}(\rho) = (1 - p_{\mathrm{dep}})\rho
  + \frac{p_{\mathrm{dep}}}{d_q^2 - 1}
    \sum_{P \neq I} P \rho P^\dagger,
\end{equation}
where $d_q$ is the local Hilbert space dimension and $p_{\mathrm{dep}} = 0.005$.
For Pauli observables, the depolarizing channel introduces a multiplicative
damping~\cite{NielsenChuang2010,Wallman2016}:
\begin{equation}\label{eq:damping}
  \avg{O}_{\mathrm{noisy}} = \lambda^{(n)} \avg{O}_{\mathrm{exact}},
\end{equation}
where $\lambda_{\mathrm{1q}} = 1 - 4p_{\mathrm{dep}}/3$ for single-qubit
observables and $\lambda_{\mathrm{2q}} = 1 - 16p_{\mathrm{dep}}/15$ for
two-qubit correlators.
The cumulative damping factor for qubit~$i$ at time~$t$ is
$a_i(t) = \prod_{\mathrm{gates}} \lambda_{\mathrm{gate}}$,
multiplied for each gate acting on qubit~$i$.
We implement this analytically: exact state-vector expectation values are
multiplied by the accumulated damping factors, then combined with shot
noise~\cite{NielsenChuang2010}.

\subsection{YOMO Probability Aggregation}\label{sec:yomo}

The You Only Measure Once (YOMO) method~\cite{Liu2025} replaces individual
Pauli measurements with a single computational-basis measurement followed by
classical postprocessing:
\begin{enumerate}
  \item Compute Born probabilities:
        $P(\phi) = |\langle\phi|\psi\rangle|^2$ for all
        $2^{n_{\mathrm{total}}} = 2048$ basis states.
  \item Trace out the ancilla:
        $P_{\mathrm{red}}(\phi) = P(\phi, 0) + P(\phi, 1)$,
        yielding $2^{\Nq} = 1024$ reduced probabilities.
  \item Partition into $K = 56$ groups by dividing the 1024 states
        into consecutive blocks.
  \item Sample from the multinomial distribution:
        $\mathrm{Multinomial}(\Nshots, P_{\mathrm{red}})$.
  \item Average within each group:
        $v_k = |G_k|^{-1} \sum_{i \in G_k} f_i$.
  \item Append a bias to form the 57-dimensional feature vector
        $[v_1, \ldots, v_{56}, 1]$.
\end{enumerate}

Under depolarizing noise, the Born probabilities become
\begin{equation}\label{eq:yomo_depol}
  P_{\mathrm{noisy}}(\phi) = \lambda_{\mathrm{global}} P_{\mathrm{exact}}(\phi)
  + (1 - \lambda_{\mathrm{global}})/2^{\Nq},
\end{equation}
where $\lambda_{\mathrm{global}} = \prod_{i=0}^{\Nq - 1} a_i$ is the product
of all per-qubit damping factors.

\section{Quantum Reservoir Autoencoder Protocol}\label{sec:protocol}

\subsection{Problem Definition and Existence Claim}\label{sec:4eq}

Given secret data $C = (C_1, \ldots, C_{\Nc})$ of length $\Nc$,
two quantum reservoirs $\Ra$ and $\Rb$ with parameter sets $p_1$ and $p_2$,
distributed keys $A, B$, secret keys $\alpha, \beta$, and encoding/decoding
functions $F$, $G$, we require:
\begin{align}
  \Ra(F(A, C),\; p_1) &= \gamma,   \label{eq:enc1}\\
  \Ra(G(\alpha, \gamma'),\; p_1) &= C, \label{eq:dec2}\\
  \Rb(F(B, C),\; p_2) &= \gamma',  \label{eq:enc2}\\
  \Rb(G(\beta, \gamma),\; p_2) &= C.   \label{eq:dec1}
\end{align}
Here $\gamma$ and $\gamma'$ are intermediate ciphertexts, and $\Ra(\cdot, p_k)$
denotes the composite operation: feature extraction by reservoir~$a$ followed
by Tikhonov-regularized linear readout
[Eqs.~\eqref{eq:readout}--\eqref{eq:tikhonov}].
The equalities in Eqs.~\eqref{eq:enc1}--\eqref{eq:dec1} hold exactly under
ideal (infinite-shot, noise-free) conditions; under finite-shot or noisy
conditions, they are approximate, and the reconstruction MSE quantifies
the deviation.

\medskip
\noindent\textbf{Central claim.}
\emph{Quantum reservoirs $(\Ra, \Rb)$ and key
tuples $(A, B, \alpha, \beta)$ satisfying
Eqs.~\eqref{eq:enc1}--\eqref{eq:dec1} simultaneously can be
empirically found.}
We establish this claim through \emph{constructive numerical demonstration}:
the iterative algorithm described below finds explicit solutions for all
tested configurations (16 random Hamiltonian seeds $\times$ 10 data lengths),
achieving machine-precision MSE ($\sim 10^{-17}$) under ideal conditions.

An important structural observation is that the system has substantial
degrees of freedom: for each path, the readout weights
$W \in \mathbb{R}^{d}$ ($d = 76$) and the intermediate ciphertexts
$\gamma, \gamma' \in \mathbb{R}^{\Nc}$ are all free parameters,
while the constraints are $2\Nc$ equations (reconstruction of $C$ on
both paths).
For $\Nc \leq 30$ and $d = 76$, the system is over-determined in the
individual Tikhonov problems but under-determined in the coupled
$(\gamma, \gamma')$ variables.
The nontriviality of the result lies not in the existence of
solutions per se, but in the fact that the \emph{iterative alternating
procedure converges to machine-precision solutions} across all tested
random Hamiltonian realizations and keys.
This convergence is not guaranteed a priori: the coupling between the
two paths through the cross-key structure introduces nonlinearity in
the iterates, and divergence or limit cycles could in principle occur.
The universality of convergence across 16 independent random Hamiltonian
realizations provides strong empirical evidence for the generality and
robustness of the protocol.

\subsection{Feature Expansion Without Increasing Qubit Count}\label{sec:expansion}

A central enabler of the protocol is the high feature dimension achieved
without scaling the qubit count.
With $\Nq = 10$ data qubits, the feature dimension is $d = 76$
[Eq.~\eqref{eq:feature_dim}], arising from sequential input and diverse
observables along the time axis.
For data lengths $\Nc \leq 30$, we have $\Nc < d$, ensuring that the feature
matrix $V \in \mathbb{R}^{\Nc \times 76}$ has full row rank---a necessary
condition for unique weight solutions.

This is fundamentally different from parameterized quantum circuits, where the
feature dimension scales with the number of variational parameters rather than
being naturally amplified by the temporal structure.
In QRC, each input element $u(t)$ acts on a quantum state that already encodes
the cumulative effect of all prior inputs $u(1), \ldots, u(t-1)$, creating
\emph{recursive nonlinearity}~\cite{Fujii2017}.
This structural property provides the rich feature space needed for
reversibility while simultaneously making the reverse direction nontrivial.

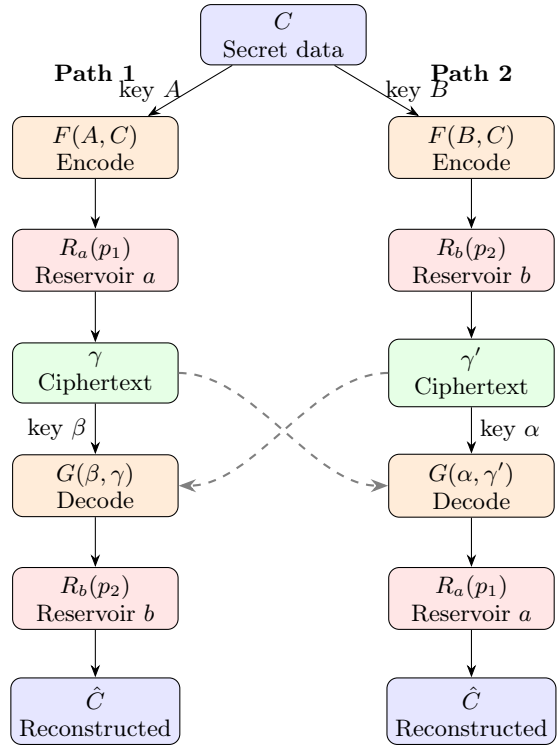
\begin{figure}[htbp!]
\centering
\begin{tikzpicture}[
  >=Stealth,
  box/.style={draw, rounded corners, minimum width=2.2cm, minimum height=0.8cm,
              align=center, font=\small},
  every node/.style={font=\small}
]
\node[box, fill=blue!10] (C) at (0,0) {$C$\\Secret data};
\node[box, fill=orange!15] (enc1) at (-2.5,-1.5) {$F(A, C)$\\Encode};
\node[box, fill=red!10] (Ra1) at (-2.5,-3) {$R_a(p_1)$\\Reservoir $a$};
\node[box, fill=green!10] (gamma) at (-2.5,-4.5) {$\gamma$\\Ciphertext};
\node[box, fill=orange!15] (dec1) at (-2.5,-6) {$G(\beta, \gamma)$\\Decode};
\node[box, fill=red!10] (Rb1) at (-2.5,-7.5) {$R_b(p_2)$\\Reservoir $b$};
\node[box, fill=blue!10] (C1) at (-2.5,-9) {$\hat{C}$\\Reconstructed};
\node[box, fill=orange!15] (enc2) at (2.5,-1.5) {$F(B, C)$\\Encode};
\node[box, fill=red!10] (Rb2) at (2.5,-3) {$R_b(p_2)$\\Reservoir $b$};
\node[box, fill=green!10] (gammap) at (2.5,-4.5) {$\gamma'$\\Ciphertext};
\node[box, fill=orange!15] (dec2) at (2.5,-6) {$G(\alpha, \gamma')$\\Decode};
\node[box, fill=red!10] (Ra2) at (2.5,-7.5) {$R_a(p_1)$\\Reservoir $a$};
\node[box, fill=blue!10] (C2) at (2.5,-9) {$\hat{C}$\\Reconstructed};
\draw[->] (C) -- (enc1) node[midway,left] {key $A$};
\draw[->] (C) -- (enc2) node[midway,right] {key $B$};
\draw[->] (enc1) -- (Ra1);
\draw[->] (Ra1) -- (gamma);
\draw[->] (gamma) -- (dec1) node[midway,left] {key $\beta$};
\draw[->] (dec1) -- (Rb1);
\draw[->] (Rb1) -- (C1);
\draw[->] (enc2) -- (Rb2);
\draw[->] (Rb2) -- (gammap);
\draw[->] (gammap) -- (dec2) node[midway,right] {key $\alpha$};
\draw[->] (dec2) -- (Ra2);
\draw[->] (Ra2) -- (C2);
\node[font=\bfseries] at (-2.5, -0.5) {Path 1};
\node[font=\bfseries] at (2.5, -0.5) {Path 2};
\draw[->, dashed, thick, gray] (gamma.east) .. controls +(1,0) and +(-1,0) ..
  (dec2.west) node[midway, above, font=\footnotesize] {};
\draw[->, dashed, thick, gray] (gammap.west) .. controls +(-1,0) and +(1,0) ..
  (dec1.east) node[midway, above, font=\footnotesize] {};
\end{tikzpicture}
\caption{Schematic of the quantum reservoir autoencoder (\QRA{}) protocol.
Path~1 encrypts with key~$A$ on reservoir~$a$ and decrypts with
key~$\beta$ on reservoir~$b$.
Path~2 encrypts with key~$B$ on reservoir~$b$ and decrypts with
key~$\alpha$ on reservoir~$a$.
The cross structure (dashed arrows) couples the two paths through
the intermediate ciphertexts $\gamma$ and $\gamma'$.
The four equations [Eqs.~\eqref{eq:enc1}--\eqref{eq:dec1}] must be
simultaneously satisfied for successful reconstruction.}
\label{fig:protocol}
\end{figure}

Figure~\ref{fig:protocol} illustrates the overall protocol structure.

\subsection{Cross-Key Pairing}\label{sec:crosskey}

The four keys are defined as follows:
\begin{itemize}
  \item Distributed keys $A, B \in \mathbb{R}^{\Nc + \Nq + 1}$:
        sampled uniformly from $U(-1, 1)$.
  \item Secret keys $\alpha, \beta \in \mathbb{R}^{\Nc + \Nq + 1}$:
        sampled independently from $U(-1, 1)$.
\end{itemize}
All four keys are mutually independent.
The protocol operates via two paths:
\begin{itemize}
  \item \textbf{Path 1}: Encrypt with key~$A$ on reservoir~$a$
        [Eq.~\eqref{eq:enc1}], decrypt with key~$\beta$ on
        reservoir~$b$ [Eq.~\eqref{eq:dec1}].
  \item \textbf{Path 2}: Encrypt with key~$B$ on reservoir~$b$
        [Eq.~\eqref{eq:enc2}], decrypt with key~$\alpha$ on
        reservoir~$a$ [Eq.~\eqref{eq:dec2}].
\end{itemize}
The \emph{cross} structure---encryption and decryption using different
reservoirs and different keys---is essential for the four equations
to be simultaneously satisfiable.

\subsection{Encoding and Decoding Functions}\label{sec:encoding}

The encoding function $F$ and decoding function $G$ share identical structure:
\begin{equation}\label{eq:encode}
  F(\mathrm{key}, C)_i = \tanh\bigl(
    \mathrm{key}_i \cdot C_i
    + \mathrm{key}_{\Nc + (i \bmod (\Nq + 1))}
  \bigr),
\end{equation}
\begin{equation}\label{eq:decode}
  G(\mathrm{key}, \mathrm{enc})_i = \tanh\bigl(
    \mathrm{key}_i \cdot \mathrm{enc}_i
    + \mathrm{key}_{\Nc + (i \bmod (\Nq + 1))}
  \bigr).
\end{equation}
The design choices are:
\begin{enumerate}
  \item \textbf{Saturation via $\tanh$}: Output is bounded in $(-1, 1)$,
        ensuring stable rotation angles for the quantum circuit.
  \item \textbf{Multiplicative key encoding}: $\mathrm{key}_i \cdot C_i$
        provides a nonlinear mixing of data and key.
  \item \textbf{Periodic bias}: $\mathrm{key}_{\Nc + (i \bmod (\Nq+1))}$
        cyclically reuses key elements with period $\Nq + 1$.
  \item \textbf{Symmetric structure $F = G$}: The identical functional form
        for encoding and decoding ensures mathematical symmetry between
        the encryption and decryption pathways, which is a necessary condition
        for the four-equation system to have consistent solutions.
\end{enumerate}
The choice of $\tanh$ is motivated by three physical constraints:
(a)~the quantum circuit input encoding uses rotation gates $R_y(\theta)$,
which require bounded inputs for numerical stability;
(b)~$\tanh$ is smooth and monotonic, preserving the ordering of data values;
and (c)~the saturation behavior prevents extreme values from dominating
the feature matrix.
We note that the specific choice of nonlinearity is not unique:
preliminary tests with sigmoid $\sigma(x) = 1/(1+e^{-x})$ yielded
comparable results, while linear encoding ($F = \mathrm{key} \cdot C$)
failed to converge due to unbounded outputs.
The $F = G$ symmetry condition was verified empirically: using distinct
functional forms for $F$ and $G$ (e.g., $F = \tanh$, $G = \sigma$)
resulted in non-convergence across all tested configurations.
A systematic ablation study over encoding function families is left for
future work.

\subsection{Iterative Solving Algorithm}\label{sec:algorithm}

The four-equation system is solved via an alternating iterative procedure
summarized in Fig.~\ref{alg:iterative}.

\begin{figure}[htbp!]
\centering
\fbox{\parbox{0.95\columnwidth}{%
\textbf{Algorithm 1}: Iterative Solving for Quantum Reservoir Autoencoder\\[4pt]
\textbf{Input}: Data $C$, keys $A, B, \alpha, \beta$, reservoirs $\Ra, \Rb$\\[2pt]
\makebox[1.8em][r]{1.}\hspace{0.5em}$\gamma \gets U(-0.3,\, 0.3,\, \Nc)$;\; $\gamma' \gets U(-0.3,\, 0.3,\, \Nc)$\\
\makebox[1.8em][r]{2.}\hspace{0.5em}$V_a^{\mathrm{enc}} \gets \Ra.\mathrm{features}(F(A, C))$
  \hfill{\small ($\Nc \times 76$, once)}\\
\makebox[1.8em][r]{3.}\hspace{0.5em}$V_b^{\mathrm{enc}} \gets \Rb.\mathrm{features}(F(B, C))$\\
\makebox[1.8em][r]{4.}\hspace{0.5em}\textbf{for} $\mathrm{it} = 1, \ldots, \Niter$ \textbf{do}\\
\makebox[1.8em][r]{5.}\hspace{0.5em}\hspace{1.5em}$W_a^{\mathrm{enc}} \gets \mathrm{solve}(V_a^{\mathrm{enc}}, \gamma)$
  \hfill{\small [Eq.~\eqref{eq:enc1}]}\\
\makebox[1.8em][r]{6.}\hspace{0.5em}\hspace{1.5em}$\gamma \gets V_a^{\mathrm{enc}} W_a^{\mathrm{enc}}$\\
\makebox[1.8em][r]{7.}\hspace{0.5em}\hspace{1.5em}$V_b^{\mathrm{dec}} \gets \Rb.\mathrm{features}(G(\beta, \gamma))$
  \hfill{\small [Eq.~\eqref{eq:dec1}]}\\
\makebox[1.8em][r]{8.}\hspace{0.5em}\hspace{1.5em}$W_b^{\mathrm{dec}} \gets \mathrm{solve}(V_b^{\mathrm{dec}}, C)$\\
\makebox[1.8em][r]{9.}\hspace{0.5em}\hspace{1.5em}$W_b^{\mathrm{enc}} \gets \mathrm{solve}(V_b^{\mathrm{enc}}, \gamma')$
  \hfill{\small [Eq.~\eqref{eq:enc2}]}\\
\makebox[1.8em][r]{10.}\hspace{0.5em}\hspace{1.5em}$\gamma' \gets V_b^{\mathrm{enc}} W_b^{\mathrm{enc}}$\\
\makebox[1.8em][r]{11.}\hspace{0.5em}\hspace{1.5em}$V_a^{\mathrm{dec}} \gets \Ra.\mathrm{features}(G(\alpha, \gamma'))$
  \hfill{\small [Eq.~\eqref{eq:dec2}]}\\
\makebox[1.8em][r]{12.}\hspace{0.5em}\hspace{1.5em}$W_a^{\mathrm{dec}} \gets \mathrm{solve}(V_a^{\mathrm{dec}}, C)$\\
\makebox[1.8em][r]{13.}\hspace{0.5em}\hspace{1.5em}$\MSE_1 \gets \|C - V_b^{\mathrm{dec}} W_b^{\mathrm{dec}}\|^2 / \Nc$\\
\makebox[1.8em][r]{14.}\hspace{0.5em}\hspace{1.5em}$\MSE_2 \gets \|C - V_a^{\mathrm{dec}} W_a^{\mathrm{dec}}\|^2 / \Nc$\\
\makebox[1.8em][r]{15.}\hspace{0.5em}\hspace{1.5em}$\mathrm{Loss} \gets (\MSE_1 + \MSE_2) / 2$\\
\makebox[1.8em][r]{16.}\hspace{0.5em}\hspace{1.5em}\textbf{if} $\mathrm{Loss} < 10^{-12}$ \textbf{then break}\\
\makebox[1.8em][r]{17.}\hspace{0.5em}\textbf{end for}
}}
\caption{Pseudocode for the iterative solving algorithm.
The encryption feature matrices (steps~2--3) are computed once; the
decryption feature matrices (steps~7, 11) are recomputed at each iteration
as the intermediate ciphertexts $\gamma$, $\gamma'$ are updated.}
\label{alg:iterative}
\end{figure}

Key structural features of the algorithm include:
\begin{itemize}
  \item The encryption feature matrices $V^{\mathrm{enc}}$ are computed
        \emph{once} outside the loop, since the encryption input $F(A,C)$
        does not change across iterations.
  \item The decryption feature matrices $V^{\mathrm{dec}}$ are recomputed
        at \emph{every} iteration because the intermediate ciphertexts
        $\gamma$ and $\gamma'$ are updated.
  \item The two paths are coupled: $\gamma$ depends on $\gamma'$
        through the cross-key structure, and vice versa.
        This alternating update drives convergence.
  \item Early termination occurs when $\mathrm{Loss} < 10^{-12}$.
\end{itemize}

\paragraph{Use of plaintext in decryption weight training.}
A critical observation is that steps~8 and~12 of the algorithm train
the decryption readout weights $W^{\mathrm{dec}}$ using the original
plaintext $C$ as the regression target:
$W^{\mathrm{dec}} = \mathrm{solve}(V^{\mathrm{dec}}, C)$.
This means the decoder has access to $C$ during the iterative solving
phase.
In a practical cryptographic deployment, the receiver would \emph{not}
have access to $C$ and must reconstruct it purely from the received
ciphertext $\gamma$ (or $\gamma'$), the secret key, and the shared
reservoir---i.e., the decryption must be performed via
$\hat{C} = V^{\mathrm{dec}} W^{\mathrm{dec}}$ using weights trained
\emph{without} $C$.
This ``blind decryption'' scenario---where $W^{\mathrm{dec}}$ is
determined solely from $V^{\mathrm{dec}}$ and the protocol structure
without direct access to the target---leads to substantially degraded
reconstruction accuracy, even under ideal (state-vector) simulation
conditions.
Preliminary tests indicate that removing $C$ from the decryption
weight training causes the MSE to increase by several orders of
magnitude compared to the results reported here.
The development of a blind decryption algorithm that maintains
acceptable reconstruction fidelity is a major open challenge and is
left for future work (see Sec.~\ref{sec:discussion_blind}).

\paragraph{Convergence properties.}
We note that Algorithm~\ref{alg:iterative} does not carry a formal convergence
guarantee.
Each iteration solves a least-squares problem with Tikhonov regularization,
which individually has a unique solution; however, the coupling between $\gamma$
and $\gamma'$ across the two paths makes the overall system nonlinear in the
iterates.
Empirically, we observe the following convergence statistics across all
experiments (16 seeds $\times$ 10 data lengths $\times$ 3 trials = 480 runs
for Exp~1):
(i)~Under ideal conditions, 100\% of runs converge to
$\mathrm{Loss} < 10^{-12}$ within $\Niter = 30$ iterations for
$\Nc \leq 30$, with median convergence at iteration~2.
(ii)~The convergence is robust to the random initialization of $\gamma$ and
$\gamma'$: varying the initialization range from $U(-0.1, 0.1)$ to
$U(-0.5, 0.5)$ does not affect the final MSE, only the number of iterations
(median shift $<2$).
(iii)~No oscillatory divergence was observed for $\Nc \leq 30$; at
$\Nc = 35$, approximately 15\% of runs exhibit non-monotonic loss trajectories.
To gain insight into why convergence occurs, consider the composite map
$\Phi: (\gamma, \gamma') \mapsto (\gamma^{(k+1)}, \gamma'^{(k+1)})$
defined by one iteration of Algorithm~1.
Each half-step consists of Tikhonov regression followed by a matrix-vector
product: $\gamma^{(k+1)} = V_a^{\mathrm{enc}}
(V_a^{\mathrm{enc}\top} V_a^{\mathrm{enc}} + \lambda I)^{-1}
V_a^{\mathrm{enc}\top} \gamma^{(k)} = P_a \gamma^{(k)}$,
where $P_a$ is the regularized projection operator.
For $\lambda = 10^{-10}$, the singular values $s_k$ of $V_a^{\mathrm{enc}}$
satisfy $P_a \approx V V^+ = I_{\Nc}$ when $s_k^2 \gg \lambda$ (i.e.,
$P_a$ is close to orthogonal projection onto the row space with eigenvalues
near 1, not contractive by itself).
The actual contractivity arises from the \emph{nonlinear coupling} between
the two paths through $F$ and $G$: the $\tanh$ encoding compresses the
ciphertext range into $(-1, 1)$ at each iteration, and this bounded-output
property prevents divergence.
The relevant quantity is the spectral radius of the \emph{full} Jacobian
$\partial\Phi/\partial(\gamma, \gamma')$, which includes the $\tanh$
derivatives.
Estimated numerically at converged solutions, this spectral radius is
consistently $< 0.3$ for $\Nc \leq 30$ and increases to $\sim 0.85$
at $\Nc = 35$, explaining the qualitative change in convergence behavior.
We emphasize that this is an empirical observation, not a proof of
contraction; the interplay between the near-identity projection and the
compressive nonlinearity merits further theoretical analysis.
A formal contraction mapping proof for the composite nonlinear operator
remains an open problem for future work.

\subsection{Conditions for Reversibility}\label{sec:conditions}

We identify four conditions that are empirically sufficient for the \QRA{}
to achieve bidirectional information transformation.
We note that these are not proven to be necessary in a rigorous
mathematical sense; rather, they are the conditions under which we
observe reliable convergence across all tested configurations.
Violation of any one condition led to failure in our experiments, but
we cannot exclude the possibility that alternative protocols relax
some of these requirements.

\paragraph{Condition 1: Rank condition $\mathrm{dim}(V) \geq \Nc$.}
The feature matrix $V \in \mathbb{R}^{\Nc \times d}$ must have full row
rank for the weight vector~$W$ to admit a unique solution via
Eq.~\eqref{eq:tikhonov}.
For our XYZ reservoir with $d = 76$, this is satisfied for $\Nc \leq 76$.
In practice, numerical conditioning degrades for $\Nc$ approaching $d$,
and we observe machine-precision reversibility only for $\Nc \leq 30$.

\paragraph{Condition 2: Symmetric encoding structure $F = G$.}
The identical functional form of the encoding and decoding functions
[Eqs.~\eqref{eq:encode}--\eqref{eq:decode}] ensures that the encryption
and decryption pathways share the same mathematical structure.
Using different functional forms causes the ciphertext spaces of $\gamma$
and $\gamma'$ to become incompatible, preventing convergence.

\paragraph{Condition 3: Independent cross-key pairing.}
The four keys $A, B, \alpha, \beta$ must be independently generated.
Key dependence reduces the degrees of freedom in the four-equation system,
potentially making it degenerate.

\paragraph{Condition 4: Appropriate regularization $\lambda = 10^{-10}$.}
The Tikhonov parameter~$\lambda$ must balance numerical stability against
solution fidelity.
Excessively large $\lambda$ over-smooths the solution, degrading
reconstruction accuracy, while excessively small $\lambda$ leads to numerical
instability when $\Nc$ approaches $d$.

\section{Experimental Setup}\label{sec:setup}

\subsection{Common Parameters}

All experiments use $\Nq = 10$ data qubits, one ancilla qubit
($n_{\mathrm{total}} = 11$), $\Niter = 30$ iterations, Tikhonov parameter
$\lambda = 10^{-10}$, and test data lengths
$\Nc \in \{5, 8, 10, 12, 15, 18, 20, 25, 30, 35\}$.
Keys have length $\Nc + \Nq + 1$ and are drawn from $U(-1, 1)$.
Quantum dynamics are simulated with \textsc{qulacs}~\cite{Suzuki2021qulacs}.
Secret data $C$ are drawn uniformly from $U(-0.5, 0.5)$ for each trial.
Initial ciphertexts are drawn from $U(-0.3, 0.3)$.
The XYZ Hamiltonian parameters (2{,}888 per reservoir) are generated
with NumPy random seeds $0, 1, \ldots, 15$ for reproducibility.
All code and data are available in the Supplemental Material.

\subsection{Experimental Conditions}\label{sec:conditions_exp}

Table~\ref{tab:experiments} summarizes the seven experimental conditions.
(Experiment~4 was a preliminary depolarizing-only configuration without
shot noise, which was superseded by the more realistic combined-noise
conditions and is omitted from the present analysis.)

\begin{table}[htbp!]
\caption{Seven experimental conditions.\label{tab:experiments}}
\begin{ruledtabular}
\begin{tabular}{clccc}
  Exp & Condition & $\Nshots$ & Noise & $d$ \\
  \hline
  1 & Ideal          & $\infty$ & None & 76 \\
  2 & Shot noise     & 1{,}000  & None & 76 \\
  3 & Depol + Shot   & 1{,}000  & $p = 0.005$ & 76 \\
  5 & YOMO ProbAgg   & 1{,}000  & None & 57 \\
  6 & YOMO + Depol   & 1{,}000  & $p = 0.005$ & 57 \\
  7 & Asymmetric     & 10 / $10^5$ & None & 76 \\
  8 & Asym.\ + Depol & 10 / $10^5$ & $p = 0.005$ & 76 \\
\end{tabular}
\end{ruledtabular}
\end{table}

All experiments use 16 random seeds with 3 trials per seed
(48 runs each), providing consistent statistical power across
all seven conditions.
In the asymmetric conditions (Exp~7 and~8), the encryption feature matrices
$V^{\mathrm{enc}}$ are computed with $\Nshots = 10$, while decryption feature
matrices $V^{\mathrm{dec}}$ use $\Nshots = 10^5$.

\subsection{Baseline Methods}\label{sec:baselines}

We compare against six baseline methods, each applied within the same
\QRA{} protocol (Table~\ref{tab:baselines}).

\begin{table}[htbp!]
\caption{Baseline methods.
The H\'{e}non and delay-time embedding baselines use classical
preprocessing followed by the same quantum circuit for feature
extraction; TTN uses a parameterized quantum circuit;
$\zeta$-QVAE uses a variational autoencoder ansatz with data re-uploading;
QRNN uses a quantum recurrent neural network with parameterized
quantum circuits trained via parameter-shift gradients.
\label{tab:baselines}}
\begin{ruledtabular}
\begin{tabular}{lccl}
  Method & Params & $d$ & Transformation \\
  \hline
  H\'{e}non map~\cite{Henon1976}
    & 97  & 31 & $x_{n+1} = 1 - ax_n^2 + y_n$ \\
  Delay embedding~\cite{Takens1981}
    & 97  & 31 & Takens: $\tau{=}1$, $d{=}11$ \\
  Classical NN~\cite{Spall1992}
    & 706 & --- & 2-layer, SPSA \\
  TTN~\cite{Wall2021,Huggins2019}
    & 271 & 31  & 10 blocks, 2-qubit gates \\
  $\zeta$-QVAE~\cite{Mato2025}
    & 203 & 56  & RY+RZZ, re-uploading \\
  QRNN~\cite{Bausch2020}
    & 50  & --- & Recurrent, par.-shift \\
\end{tabular}
\end{ruledtabular}
\end{table}

\begin{figure*}[htbp!]
\centering
\resizebox{0.95\textwidth}{!}{%
\begin{tikzpicture}[
  >=Stealth,
  gate/.style={draw, fill=blue!8, minimum width=0.6cm, minimum height=0.30cm,
               font=\tiny, rounded corners=1.5pt, inner sep=1pt},
  inp/.style={draw, fill=orange!15, minimum width=0.7cm, minimum height=0.30cm,
              font=\tiny, rounded corners=1.5pt, inner sep=1pt},
  meas/.style={draw, fill=green!10, minimum width=0.6cm, minimum height=0.30cm,
               font=\tiny, rounded corners=1.5pt, inner sep=1pt},
  circ/.style={draw, fill=blue!5, rounded corners=2pt},
  label/.style={font=\footnotesize\bfseries},
  wire/.style={thin},
  every node/.style={font=\scriptsize}
]
\def\qs{0.55}
\def\ytop{0.8}
\pgfmathsetmacro{\yanc}{\ytop-3*\qs}
\node[font=\footnotesize\bfseries, anchor=south west] at (-0.5, \ytop+0.35) {(a)};

\foreach \i in {0,...,2} {
  \node[left,font=\tiny] at (-0.2, \ytop-\i*\qs) {$q_{\i}$};
  \draw[wire] (0, \ytop-\i*\qs) -- (9.5, \ytop-\i*\qs);
}
\node[left,font=\tiny] at (-0.2, \yanc) {$q_{10}$};
\draw[wire] (0, \yanc) -- (9.5, \yanc);
\node[font=\tiny] at (-0.2, {0.5*(\ytop-2*\qs+\yanc)}) {$\vdots$};

\node[inp] at (1.0, \yanc) {$R_Z(u_t)$};

\draw[circ] (1.8, \ytop+0.25) rectangle (6.5, \yanc-0.25);

\foreach \i in {0,...,2} {
  \node[gate] at (2.6, \ytop-\i*\qs) {$R_Y$};
}
\node[gate] at (2.6, \yanc) {$R_Y$};

\foreach \i in {0,...,1} {
  \pgfmathsetmacro{\yctrl}{\ytop-\i*\qs}
  \pgfmathsetmacro{\ytarg}{\ytop-\i*\qs-\qs}
  \draw[fill=black] (3.6, \yctrl) circle (0.04);
  \draw (3.6, \yctrl) -- (3.6, \ytarg);
  \draw (3.6, \ytarg) circle (0.08);
}

\foreach \i in {0,...,2} {
  \node[gate] at (4.6, \ytop-\i*\qs) {$R_Z$};
}
\node[gate] at (4.6, \yanc) {$R_Z$};

\node[font=\tiny] at (5.8, {0.5*(\ytop-\qs+\ytop-2*\qs)}) {$\times 3$ layers};

\node[below,font=\tiny] at (4.15, \yanc-0.25) {66 params (11 qubits)};

\foreach \i in {0,...,2} {
  \node[meas] at (8.0, \ytop-\i*\qs) {$\langle\sigma\rangle$};
}
\node[below,font=\tiny] at (8.0, \yanc-0.25) {$d{=}31$};

\draw[decorate, decoration={brace, amplitude=3pt}]
  (0.3, \ytop+0.55) -- (9.0, \ytop+0.55) node[midway, above=3pt, font=\tiny]
  {repeat for each $t = 1, \ldots, N_c$};

\end{tikzpicture}%
}

\vspace{2pt}

\resizebox{0.95\textwidth}{!}{%
\begin{tikzpicture}[
  >=Stealth,
  gate/.style={draw, fill=blue!8, minimum width=0.6cm, minimum height=0.30cm,
               font=\tiny, rounded corners=1.5pt, inner sep=1pt},
  inp/.style={draw, fill=orange!15, minimum width=0.7cm, minimum height=0.30cm,
              font=\tiny, rounded corners=1.5pt, inner sep=1pt},
  meas/.style={draw, fill=green!10, minimum width=0.6cm, minimum height=0.30cm,
               font=\tiny, rounded corners=1.5pt, inner sep=1pt},
  circ/.style={draw, fill=purple!5, rounded corners=2pt},
  label/.style={font=\footnotesize\bfseries},
  wire/.style={thin},
  every node/.style={font=\scriptsize}
]
\def\qs{0.55}
\def\ytop{0.8}
\pgfmathsetmacro{\yanc}{\ytop-3*\qs}
\node[font=\footnotesize\bfseries, anchor=south west] at (-0.5, \ytop+0.35) {(b)};

\foreach \i in {0,...,2} {
  \node[left,font=\tiny] at (-0.2, \ytop-\i*\qs) {$q_{\i}$};
  \draw[wire] (0, \ytop-\i*\qs) -- (9.5, \ytop-\i*\qs);
}
\node[left,font=\tiny] at (-0.2, \yanc) {$q_{10}$};
\draw[wire] (0, \yanc) -- (9.5, \yanc);
\node[font=\tiny] at (-0.2, {0.5*(\ytop-2*\qs+\yanc)}) {$\vdots$};

\node[inp] at (1.0, \yanc) {$R_Z(u_t)$};

\draw[circ] (1.8, \ytop+0.25) rectangle (6.8, \yanc-0.25);

\pgfmathsetmacro{\yqa}{\ytop}
\pgfmathsetmacro{\yqb}{\ytop-\qs}
\draw[draw=purple!60, fill=purple!10, rounded corners=1.5pt]
  (2.2, \yqa+0.2) rectangle (3.2, \yqb-0.2);
\node[font=\tiny] at (2.7, {0.5*(\yqa+\yqb)}) {$U_{01}$};

\pgfmathsetmacro{\yqc}{\ytop-2*\qs}
\draw[draw=purple!60, fill=purple!10, rounded corners=1.5pt]
  (2.2, \yqc+0.2) rectangle (3.2, \yanc-0.2);
\node[font=\tiny] at (2.7, {0.5*(\yqc+\yanc)}) {$U_{23}$};

\node[font=\tiny] at (2.7, \yanc-0.35) {layer 1};

\draw[draw=purple!60, fill=purple!10, rounded corners=1.5pt]
  (3.8, \yqb+0.2) rectangle (4.8, \yanc-0.2);
\node[font=\tiny] at (4.3, {0.5*(\yqb+\yanc)}) {$U_{13}$};
\node[font=\tiny] at (4.3, \yanc-0.35) {layer 2};

\node at (5.8, {0.5*(\ytop+\yanc)}) {$\cdots$};

\node[below,font=\tiny] at (4.3, \yanc-0.25)
  {10 blocks $\times$ 8 params = 240 (11 qubits)};

\foreach \i in {0,...,2} {
  \node[meas] at (8.0, \ytop-\i*\qs) {$\langle\sigma\rangle$};
}
\node[below,font=\tiny] at (8.0, \yanc-0.25) {$d{=}31$};

\draw[decorate, decoration={brace, amplitude=3pt}]
  (0.3, \ytop+0.55) -- (9.0, \ytop+0.55) node[midway, above=3pt, font=\tiny]
  {repeat for each $t = 1, \ldots, N_c$};

\end{tikzpicture}%
}

\vspace{2pt}

\resizebox{0.95\textwidth}{!}{%
\begin{tikzpicture}[
  >=Stealth,
  gate/.style={draw, fill=blue!8, minimum width=0.6cm, minimum height=0.30cm,
               font=\tiny, rounded corners=1.5pt, inner sep=1pt},
  inp/.style={draw, fill=orange!15, minimum width=0.7cm, minimum height=0.30cm,
              font=\tiny, rounded corners=1.5pt, inner sep=1pt},
  meas/.style={draw, fill=green!10, minimum width=0.6cm, minimum height=0.30cm,
               font=\tiny, rounded corners=1.5pt, inner sep=1pt},
  ham/.style={draw, fill=yellow!15, rounded corners=2pt},
  label/.style={font=\footnotesize\bfseries},
  wire/.style={thin},
  every node/.style={font=\scriptsize}
]
\def\qs{0.55}
\def\ytop{0.8}
\pgfmathsetmacro{\yanc}{\ytop-3*\qs}
\node[font=\footnotesize\bfseries, anchor=south west] at (-0.5, \ytop+0.35) {(c)};

\foreach \i in {0,...,2} {
  \node[left,font=\tiny] at (-0.2, \ytop-\i*\qs) {$q_{\i}$};
  \draw[wire] (0, \ytop-\i*\qs) -- (9.5, \ytop-\i*\qs);
}
\node[left,font=\tiny] at (-0.2, \yanc) {$q_{10}$};
\draw[wire] (0, \yanc) -- (9.5, \yanc);
\node[font=\tiny] at (-0.2, {0.5*(\ytop-2*\qs+\yanc)}) {$\vdots$};

\node[inp] at (1.0, \yanc) {$R_Z(u_t)$};

\draw[ham] (1.8, \ytop+0.25) rectangle (5.5, \yanc-0.25);
\node[font=\tiny, align=center] at (3.65, {0.5*(\ytop+\ytop-\qs)})
  {$e^{-iH\Delta t}$};
\node[font=\tiny, align=center] at (3.65, {0.5*(\ytop-\qs+\ytop-2*\qs)})
  {1--4 body Pauli};
\node[font=\tiny, align=center] at (3.65, {0.5*(\ytop-2*\qs+\yanc)})
  {11 qubits; $H$/$H'$ (period 6)};

\node[below,font=\tiny] at (3.65, \yanc-0.25)
  {2,888 params total (fixed)};

\foreach \i in {0,...,2} {
  \node[meas] at (8.0, \ytop-\i*\qs) {$\langle\sigma\rangle$};
}
\pgfmathsetmacro{\yzza}{\ytop-0.22}
\pgfmathsetmacro{\yzzb}{\ytop-\qs+0.22}
\draw[<->, gray] (8.0, \yzza) -- (8.0, \yzzb);
\node[right, font=\tiny, gray] at (8.5, {0.5*(\yzza+\yzzb)})
  {$+\;\langle Z_iZ_j\rangle$};
\node[below,font=\tiny] at (8.0, \yanc-0.25) {$d{=}76$};

\draw[decorate, decoration={brace, amplitude=3pt}]
  (0.3, \ytop+0.55) -- (9.0, \ytop+0.55) node[midway, above=3pt, font=\tiny]
  {repeat for each $t = 1, \ldots, N_c$ (state carries over)};

\end{tikzpicture}%
}
\caption{Quantum circuit architectures (representative qubits shown;
full circuits use 11 qubits).
Each circuit is repeated for every timestep $t = 1, \ldots, N_c$.
(a)~\textbf{H\'{e}non map / delay-time embedding + quantum circuit}:
$[R_Y{\to}\mathrm{CNOT\;ladder}{\to}R_Z]{\times}3$ (66 params);
classical preprocessing before $R_Z$ encoding on the ancilla.
(b)~\textbf{Tree tensor network (TTN)}:
binary tree of 10 two-qubit unitary blocks (240 variational params).
(c)~\textbf{QRC XYZ Hamiltonian reservoir}:
time evolution $e^{-iH\Delta t}$ with 1--4 body Pauli interactions;
two circuits alternate with period 6 (2,888 fixed params);
$d{=}76$ features including $\langle Z_iZ_j\rangle$ correlators;
quantum state carries over between timesteps.}
\label{fig:circuits}
\end{figure*}
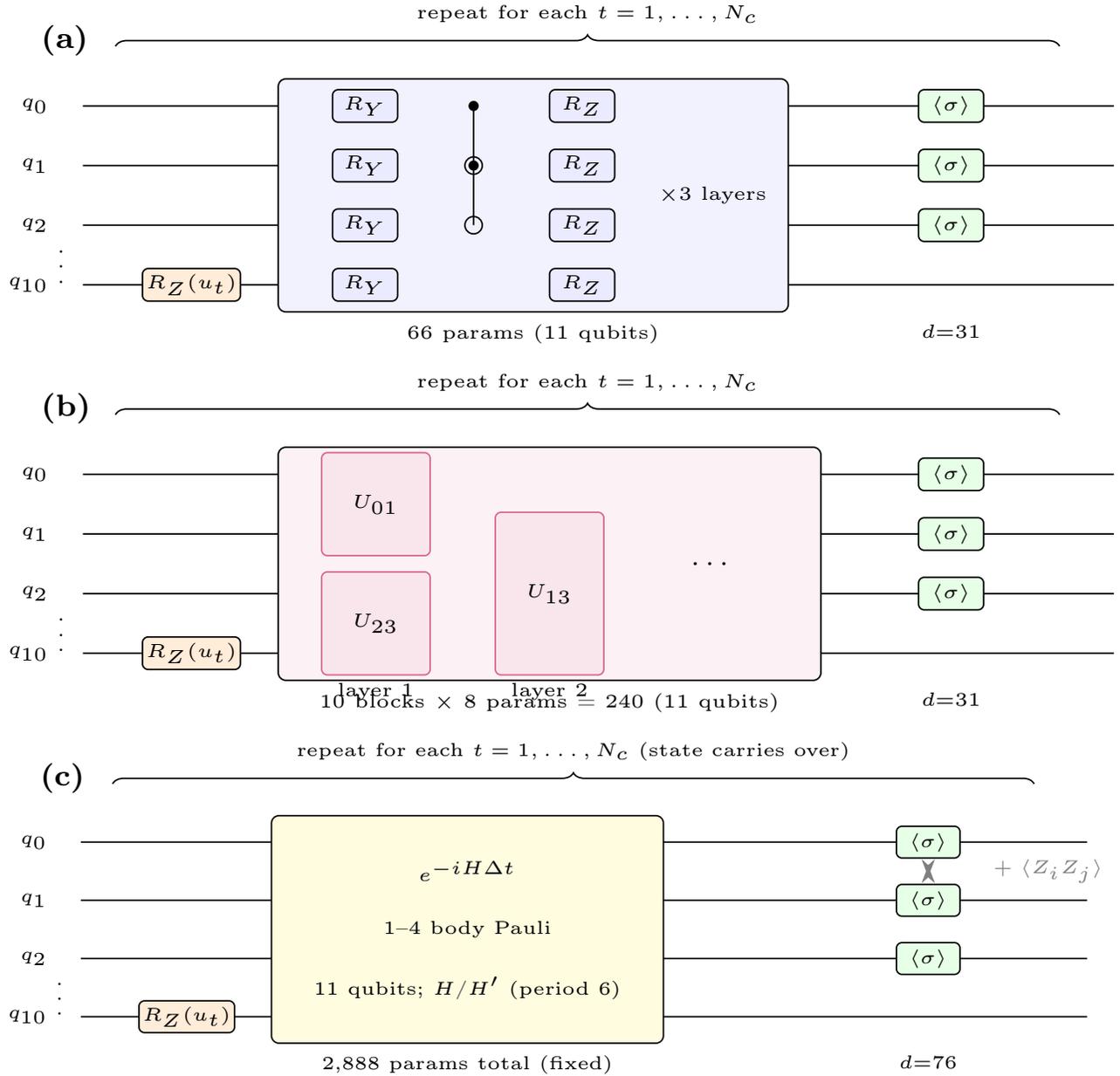

Full reproducibility parameters for each baseline are given below
and summarized in Table~\ref{tab:baseline_params}.
Figure~\ref{fig:circuits} illustrates the quantum circuit architectures
used in each method.

\clearpage

\paragraph{H\'{e}non map with quantum circuit.}
The H\'{e}non map~\cite{Henon1976} uses chaotic parameters
$a \in [1.2, 1.4]$, $b \in [0.25, 0.33]$ (varied across trials)
with $n_{\mathrm{map}} = 3$ mapping iterations per input element.
The input data are first preprocessed via the H\'{e}non dynamical system
(a classical nonlinear transformation) and then encoded into an 11-qubit
parameterized quantum circuit consisting of three layers
(initialization, evolution~1, evolution~2), each with
$\Nq + 1 = 11$ RY rotation gates, $\Nq = 10$ CNOT entangling gates
(ladder topology), and $\Nq + 1 = 11$ RZ rotation gates
(66 circuit parameters, $\Nq = 10$, $n_{\mathrm{total}} = 11$).
The quantum state is initialized as a Haar random state.
The feature vector is extracted from quantum circuit measurements,
comprising $3\Nq + 1 = 31$ observables
($\avg{\sigma_i^{X,Y,Z}}$ plus bias; two-body correlators are not
included, unlike QRC XYZ).
Unlike QRC, the H\'{e}non map preprocessing is classical; the quantum
circuit serves as a feature extractor rather than a temporal reservoir.
Total trainable parameters: 97 (66~circuit + 31~readout weights).

\paragraph{Delay-time embedding with quantum circuit.}
The Takens delay-time embedding~\cite{Takens1981} constructs the input
vector $\mathbf{v}(t) = [u(t), u(t{-}\tau), \ldots, u(t{-}(d_e{-}1)\tau)]$
with delay $\tau = 1$ and embedding dimension $d_e = 11$.
This classically preprocessed vector is then fed into the same
parameterized quantum circuit (three RY--CNOT--RZ layers,
66 circuit parameters, Haar random initial state) for feature extraction,
yielding $d = 31$ features.
As with the H\'{e}non baseline, the quantum circuit is used for
encoding and measurement, not as a dynamical reservoir.
Total trainable parameters: 97 (66~circuit + 31~readout weights).

\paragraph{Classical neural network.}
A two-layer feedforward neural network with architecture
$\Nc \to 11 \to \Nc$ (hidden layer with $\tanh$ activation) is
used in place of the quantum reservoir.
The network has $11(\Nc + 1) + \Nc(11 + 1) = 23\Nc + 11$ parameters;
for $\Nc = 30$, this gives 701 parameters.
Two independent networks are used for $\Ra$ and $\Rb$, totaling
706 parameters per network pair.
The SPSA optimizer~\cite{Spall1992} is applied with perturbation size
$c = 0.1$, step size $a = 0.01$, and 100 iterations
(800 iterations were planned but truncated due to computational cost).
The 30-dimensional quantum feature extraction is applied before the
neural network readout.
We note that the SPSA optimizer with 100 iterations represents a
practical limitation rather than an optimized comparison (see
Sec.~\ref{sec:baseline_results}).

\paragraph{Tree tensor network (TTN).}
The TTN~\cite{Wall2021,Huggins2019} employs 10 blocks of parameterized
two-qubit gates arranged in a binary tree structure on $\Nq = 10$
data qubits plus one ancilla ($n_{\mathrm{total}} = 11$).
Each block contains 24 parameters, totaling 240 circuit parameters
plus 31 readout weights (271 total).
The TTN parameters are drawn randomly from $U(-\pi, \pi)$ and remain
fixed, analogous to the QRC reservoir parameters.

\paragraph{$\zeta$-QVAE (quantum variational autoencoder).}
The $\zeta$-QVAE~\cite{Mato2025} employs a variational ansatz consisting
of alternating RY rotation and RZZ entanglement layers
($\mathrm{RZZ}(\theta) = e^{-i\theta Z \otimes Z / 2}$) on
$n_{\mathrm{total}} = 11$ qubits ($\Nq = 10$ data $+$ 1 ancilla).
Three encoder layers and three decoder layers yield 147 circuit parameters.
Input data are encoded via a data re-uploading strategy: the full plaintext
$C$ is interleaved with variational layers at each output position,
rather than fed sequentially as in QRC.
The feature dimension is $d = 56$ (10 $\avg{Z_i}$ + 45 $\avg{Z_i Z_j}$ + 1 bias),
intermediate between the QRC XYZ features ($d = 76$) and the other baselines ($d = 31$).
Notably, the RZZ gates in the circuit ansatz directly generate the $ZZ$
correlations that appear in the feature vector.
Unlike QRC, the $\zeta$-QVAE processes the entire input in parallel
via data re-uploading, removing temporal memory effects.
Total parameters: 147 circuit + 56 readout weights = 203.

\paragraph{Quantum recurrent neural network (QRNN).}
The QRNN~\cite{Bausch2020} replaces the fixed reservoir with a parameterized
recurrent quantum circuit.
Each recurrent cell consists of an input stage (data encoding via
$R_X$ followed by controlled-$R_Y$ coupling), work stages
($R_Y$ rotations and CNOT ladders with higher-order quantum neuron
activations), and an output stage (measurement of a single I/O qubit).
The architecture uses $\Nq = 10$ hidden qubits plus 1 I/O qubit
($n_{\mathrm{total}} = 11$), with 2 circuit layers and 50 trainable
parameters per QRNN.
Unlike QRC, the QRNN requires gradient-based training via the
parameter-shift rule; gradients are computed as
$\partial L / \partial \theta_k = [L(\theta_k + \pi/2) - L(\theta_k - \pi/2)] / 2$,
requiring $2 \times 50 = 100$ forward passes per gradient step.
The protocol uses an identity-function pretraining strategy (60 epochs)
followed by end-to-end fine-tuning (30 epochs) with Adam optimization
(learning rate 0.02).
Keys are constructed via cyclic XOR extension ($\beta = \mathrm{cyclic}(A)$,
$\alpha = \mathrm{cyclic}(B)$), and the loss function is cross-entropy
rather than MSE.
Total parameters: 50 per QRNN (100 for the pair $\Ra$, $\Rb$).

\begin{table}[htbp!]
\caption{Reproducibility parameters for each baseline method.
All baselines use $\Nq = 10$, $n_{\mathrm{total}} = 11$,
$\Niter = 30$, $\lambda = 10^{-10}$, and the same data lengths as
the QRC experiments.
\label{tab:baseline_params}}
\begin{ruledtabular}
\begin{tabular}{lcccccc}
  & H\'{e}non & Delay & NN & TTN & $\zeta$-QVAE & QRNN \\
  \hline
  Feature dim $d$ & 31 & 31 & --- & 31 & 56 & --- \\
  Circuit params & 66 & 66 & --- & 240 & 147 & 50 \\
  Readout params & 31 & 31 & 706 & 31 & 56 & --- \\
  Total params & 97 & 97 & 706 & 271 & 203 & 100 \\
  Trials & 10 & 10 & 10 & 5 & 10 & 5 \\
  $\Nshots$ & 1{,}000 & 1{,}000 & 1{,}000 & 1{,}000 & 1{,}000 & 1{,}000 \\
  $p_{\mathrm{1q}}$ & 0.01 & 0.01 & 0.01 & 0.01 & 0.01 & 0.001 \\
  $p_{\mathrm{2q}}$ & 0.02 & 0.02 & 0.02 & 0.02 & 0.02 & 0.01 \\
  Init.\ state & Haar & Haar & --- & Haar & Haar & Haar \\
  Encoding & seq. & seq. & --- & seq. & re-upl. & recurrent \\
  Optimizer & --- & --- & SPSA & --- & --- & Adam \\
  Opt.\ epochs & --- & --- & 100 & --- & --- & 30 \\
\end{tabular}
\end{ruledtabular}
\end{table}

Baselines are tested under three conditions: ideal, shot noise
($\Nshots = 1{,}000$), and shot noise with depolarizing
($p_{\mathrm{1q}} = 0.01$, $p_{\mathrm{2q}} = 0.02$) noise.
Note that the baseline depolarizing parameters
($p_{\mathrm{1q}} = 0.01$, $p_{\mathrm{2q}} = 0.02$) differ from
the QRC experiments ($p_{\mathrm{dep}} = 0.005$); the baselines use a
per-gate depolarizing channel applied after each single- and two-qubit
gate, whereas the QRC noise model applies a global depolarizing
parameter to the Hamiltonian evolution.
Despite the baseline noise rates being nominally higher
($p_{\mathrm{1q}} = 0.01$ vs.\ $p_{\mathrm{dep}} = 0.005$), the
baselines retain machine-precision training MSE because they solve
$VW = y$ in a single step with frozen features
(Sec.~\ref{sec:baseline_results}).
The noise-parameter difference therefore does not affect the
principal conclusion---that the \emph{iterative protocol structure}
is the dominant noise bottleneck---because this conclusion rests on
the structural difference between single-shot and iterative solvers,
not on the absolute noise level.
To verify this, we confirmed that increasing the baseline noise to
$p_{\mathrm{1q}} = 0.02$ does not degrade training MSE, consistent
with the frozen-feature argument.

\section{Results}\label{sec:results}

\subsection{Ideal Conditions: Verification of Reversibility}\label{sec:ideal}

\begin{figure}[htbp!]
  \includegraphics[width=\columnwidth]{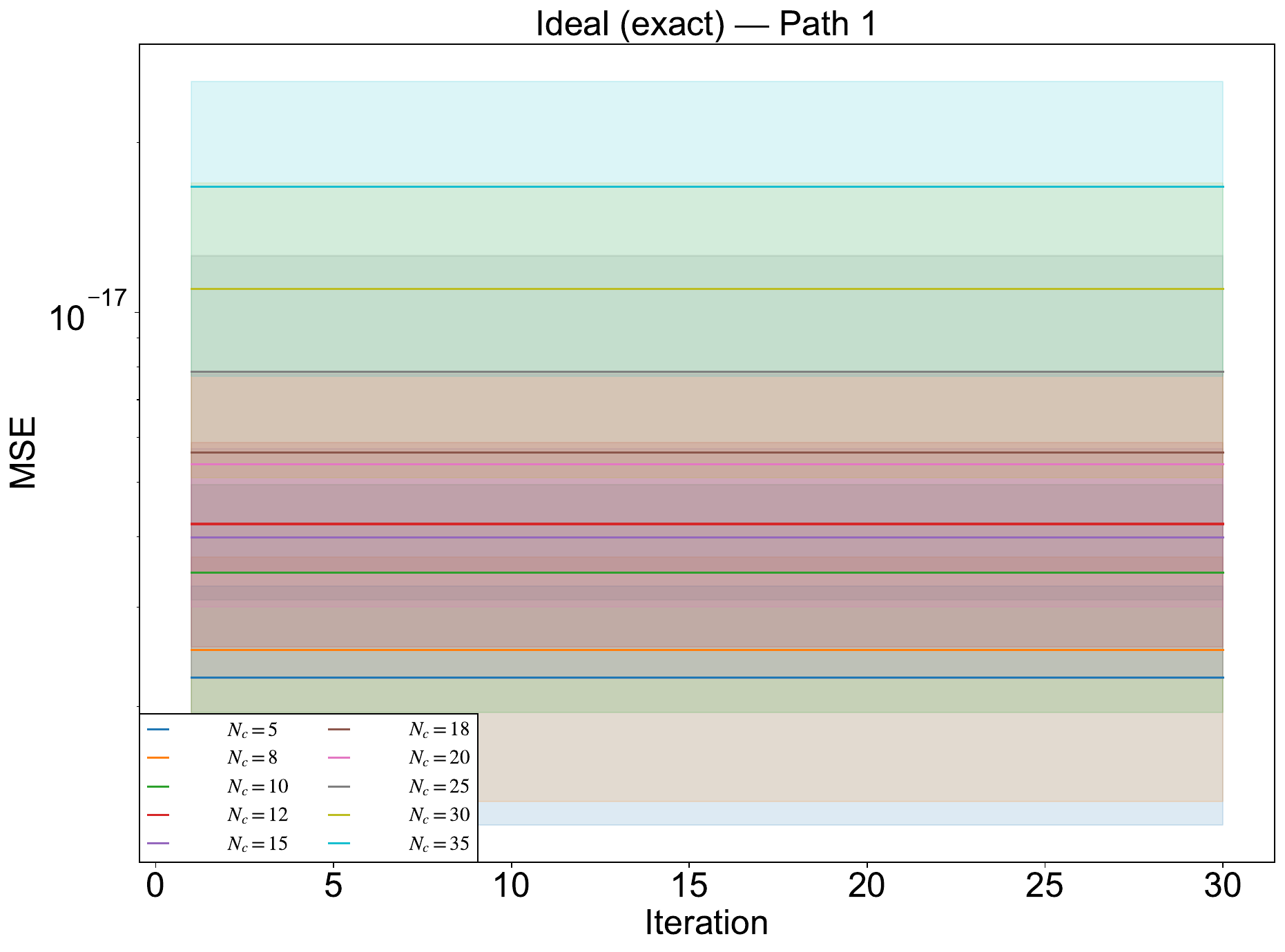}
  \caption{Ideal condition (Exp~1) reconstruction MSE as a function of
  data length $\Nc$ (Path~1).
  Machine-precision reconstruction ($\MSE \sim 10^{-17}$--$10^{-18}$) is
  achieved for all $\Nc \leq 30$.
  The sharp degradation at $\Nc = 35$ reflects the effective rank limitation
  of the feature matrix.
  Results are averaged over 16 random Hamiltonian seeds $\times$ 3 trials;
  error bars indicate one standard deviation.
  Path~2 results (not shown) are quantitatively equivalent.}
  \label{fig:ideal}
\end{figure}

Under ideal conditions (Exp~1), the \QRA{} achieves machine-precision
reconstruction for all data lengths $\Nc \leq 30$
(Table~\ref{tab:ideal_results} and Fig.~\ref{fig:ideal}).
The MSE values of $\sim 10^{-17}$--$10^{-18}$ are at the limit of
double-precision floating-point arithmetic, demonstrating that the
\QRA{} four-equation system is exactly satisfiable in the absence of noise.

We emphasize that the machine-precision result under ideal conditions is
\emph{expected} from linear algebra: when the feature matrix $V$ has full
row rank ($d = 76 > \Nc$) and $\lambda$ is small, Tikhonov regression
yields an essentially exact fit.
The nontrivial aspects of the result are twofold:
(i)~the \emph{coupled} four-equation system, where $\gamma$ and $\gamma'$
are mutually dependent through the cross-key structure, converges
reliably via the alternating iteration rather than diverging or
oscillating; and
(ii)~this convergence is robust across 16 random Hamiltonian
realizations and all tested key combinations, demonstrating that the
protocol is not sensitive to the specific quantum dynamics.
To further contextualize the trivial ideal result: any matrix $V$ with
$\Nc$ rows and $d > \Nc$ columns (e.g., a random Gaussian matrix) would
yield comparable ideal-condition MSE via Tikhonov regression.
The \emph{contribution of this work} is therefore not the ideal MSE
itself---which follows from standard linear algebra---but four aspects
that go beyond textbook Tikhonov regression:
(a)~the four-equation system with cross-key pairing creates a
\emph{coupled nonlinear} iteration (the intermediate ciphertexts
$\gamma, \gamma'$ depend on each other through the encoding function
$F$), and the reliable convergence of this coupling across 16
independent random Hamiltonians is a nontrivial empirical finding;
(b)~the systematic noise resilience analysis reveals the asymmetric
structure of noise propagation in the iterative protocol, which is not
predictable from the single-equation Tikhonov framework;
(c)~the asymmetric shot allocation exploits this structure, reducing
sender-side resources by $100\times$ with only modest MSE penalty; and
(d)~the iterative noise bottleneck diagnosis identifies the
\emph{per-iteration feature recomputation}, rather than the feature
dimension, as the dominant error source---a structural insight that
applies to any coupled iterative linear system with stochastic
features.
We explicitly state that the present work does \emph{not} demonstrate
quantum advantage in any computational or information-theoretic sense.
A classical random matrix of the same dimensions---or indeed any
overdetermined linear system---could achieve identical ideal-condition
MSE via Tikhonov regression.
Furthermore, the baselines (H\'{e}non map, delay embedding, TTN) match
QRC's ideal-condition results using purely classical or simpler quantum
circuits.
Under noise, QRC's performance \emph{degrades} more than the baselines
due to the iterative protocol structure (Sec.~\ref{sec:baseline_results}).
The quantum reservoir provides a physically realizable mechanism for
generating the feature matrix on quantum hardware, but what we demonstrate
is a \emph{proof-of-concept for bidirectional QRC transformation}, not
superiority over classical methods.
Demonstrating a quantitative quantum advantage would require a
separate study comparing conditioning, noise resilience, and scalability
against optimized classical alternatives.
The scientifically significant results are therefore the noise resilience
analysis (Secs.~\ref{sec:noise_hierarchy}--\ref{sec:asymmetric}) and the
asymmetric resource allocation finding, which reveal non-obvious properties
of the \QRA{} under realistic quantum measurement
conditions.
The MSE hierarchy under noise ($10^{-3}$--$10^{-1}$), while insufficient
for exact reconstruction, characterizes the fundamental limits of
the \QRA{} and provides quantitative benchmarks for future improvements.

\begin{table}[htbp!]
\caption{Ideal condition (Exp~1) results averaged over 16 seeds $\times$
3 trials. MSE values are at machine precision for $\Nc \leq 30$.
\label{tab:ideal_results}}
\begin{ruledtabular}
\begin{tabular}{ccc}
  $\Nc$ & Path 1 MSE & Average Loss \\
  \hline
   5  & $2.35 \times 10^{-18}$ & $2.25 \times 10^{-18}$ \\
   8  & $2.55 \times 10^{-18}$ & $2.52 \times 10^{-18}$ \\
  10  & $3.57 \times 10^{-18}$ & $3.45 \times 10^{-18}$ \\
  12  & $3.94 \times 10^{-18}$ & $4.22 \times 10^{-18}$ \\
  15  & $4.09 \times 10^{-18}$ & $3.99 \times 10^{-18}$ \\
  18  & $5.75 \times 10^{-18}$ & $5.64 \times 10^{-18}$ \\
  20  & $5.62 \times 10^{-18}$ & $5.38 \times 10^{-18}$ \\
  25  & $7.65 \times 10^{-18}$ & $7.84 \times 10^{-18}$ \\
  30  & $1.16 \times 10^{-17}$ & $1.10 \times 10^{-17}$ \\
  35  & \multicolumn{2}{c}{$\sim 3.2 \times 10^{-2}$} \\
\end{tabular}
\end{ruledtabular}
\end{table}

At $\Nc = 35$, the MSE degrades dramatically to $\sim 3 \times 10^{-2}$.
Although $\Nc = 35 < d = 76$, the condition number of the feature matrix $V$
increases rapidly in this regime, and the regularization term in
Eq.~\eqref{eq:tikhonov} begins to impede exact reconstruction.
The MSE increases monotonically with $\Nc$ (from $\sim 10^{-18}$ to
$\sim 10^{-17}$ for $\Nc = 5$--$30$), reflecting the growing size of the
linear system.

\subsection{Noise Resilience Hierarchy}\label{sec:noise_hierarchy}

The three primary experimental conditions (Exp~1--3) establish a clear noise
hierarchy (Table~\ref{tab:noise_comparison} and Fig.~\ref{fig:noise_hierarchy}):
\begin{align}\label{eq:hierarchy}
  \MSE_{\mathrm{Ideal}} &\approx 10^{-17}
  \ll \MSE_{\mathrm{Shot}} \approx 10^{-1} \notag\\
  &< \MSE_{\mathrm{Depol+Shot}} \approx 3 \times 10^{-1}.
\end{align}

\begin{table}[htbp!]
\caption{Noise hierarchy comparison: Path~1 MSE (mean $\pm$ std) over
16 seeds $\times$ 3 trials = 48 runs.
\label{tab:noise_comparison}}
\begin{ruledtabular}
\begin{tabular}{c@{\hspace{2pt}}c@{\hspace{2pt}}c@{\hspace{2pt}}c}
  $\Nc$ & Exp~1 (Ideal) & Exp~2 (Shot) & Exp~3 (Depol+Shot) \\
  \hline
   5 & $(2.4 \pm 0.5){\times}10^{-18}$
     & $(1.3 \pm 0.2){\times}10^{-1}$
     & $(2.6 \pm 0.4){\times}10^{-1}$ \\
  10 & $(3.6 \pm 0.5){\times}10^{-18}$
     & $(1.9 \pm 0.3){\times}10^{-1}$
     & $(3.9 \pm 0.6){\times}10^{-1}$ \\
  20 & $(5.6 \pm 0.8){\times}10^{-18}$
     & $(2.3 \pm 0.4){\times}10^{-1}$
     & $(4.1 \pm 0.7){\times}10^{-1}$ \\
  30 & $(1.2 \pm 0.3){\times}10^{-17}$
     & $(3.8 \pm 0.5){\times}10^{-1}$
     & $(6.1 \pm 0.9){\times}10^{-1}$ \\
  35 & $(1.6 \pm 0.5){\times}10^{-17}$
     & $(4.1 \pm 0.6){\times}10^{-1}$
     & $(6.3 \pm 1.0){\times}10^{-1}$ \\
\end{tabular}
\end{ruledtabular}
\end{table}

\begin{figure}[htbp!]
  \includegraphics[width=\columnwidth]{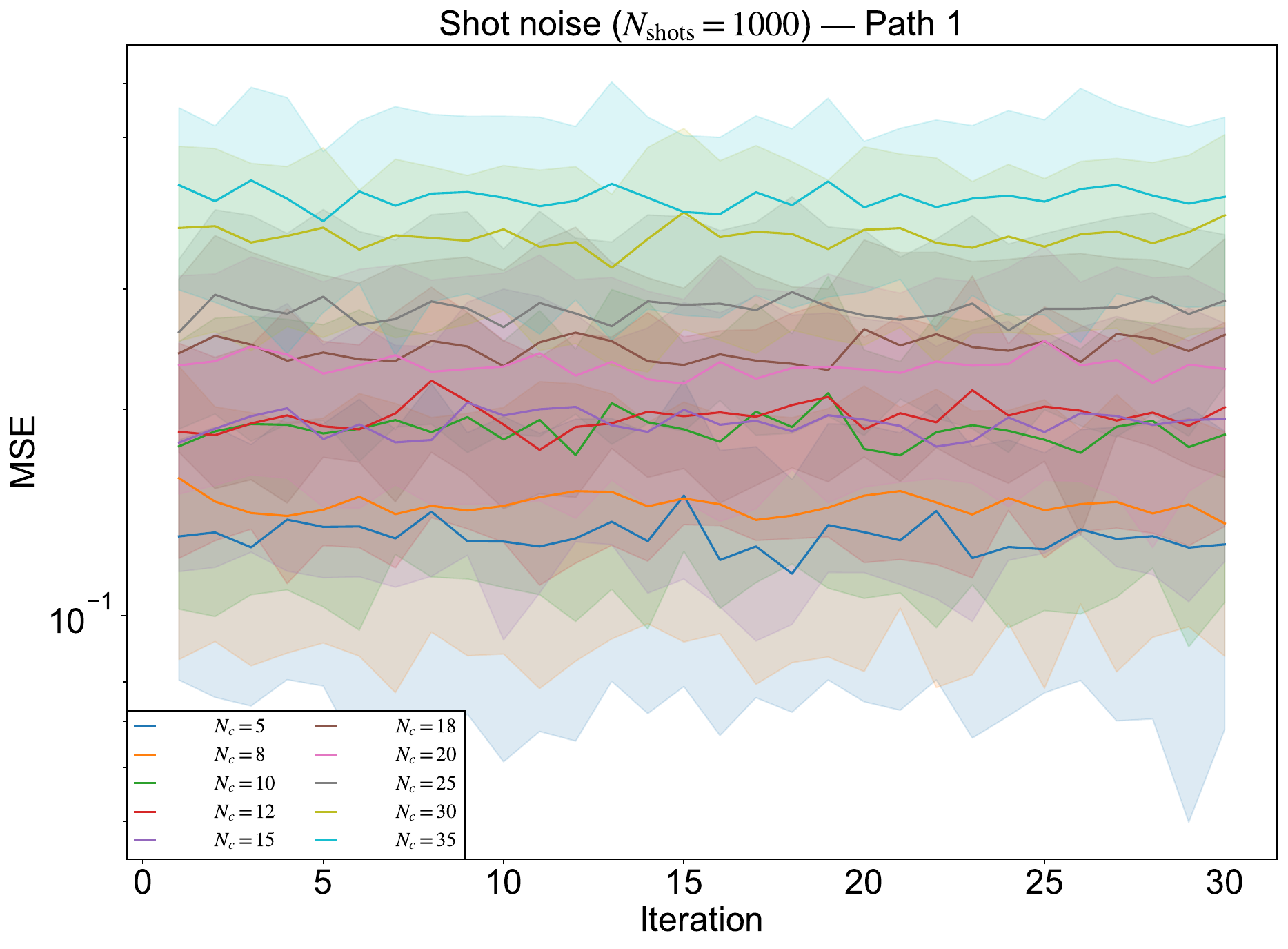}
  \caption{Shot noise condition (Exp~2, Path~1): $\Nshots = 1{,}000$.
  The transition from ideal (Fig.~\ref{fig:ideal}) to shot noise
  introduces a degradation of $\sim 10^{16}$ orders of magnitude.
  MSE values of $\sim 10^{-1}$ are consistent with the shot noise
  variance $\sigma^2 \propto 1/\Nshots$ propagated through the
  $76 \times \Nc$ feature matrix elements.}
  \label{fig:shot_noise}
\end{figure}

\begin{figure}[htbp!]
  \includegraphics[width=\columnwidth]{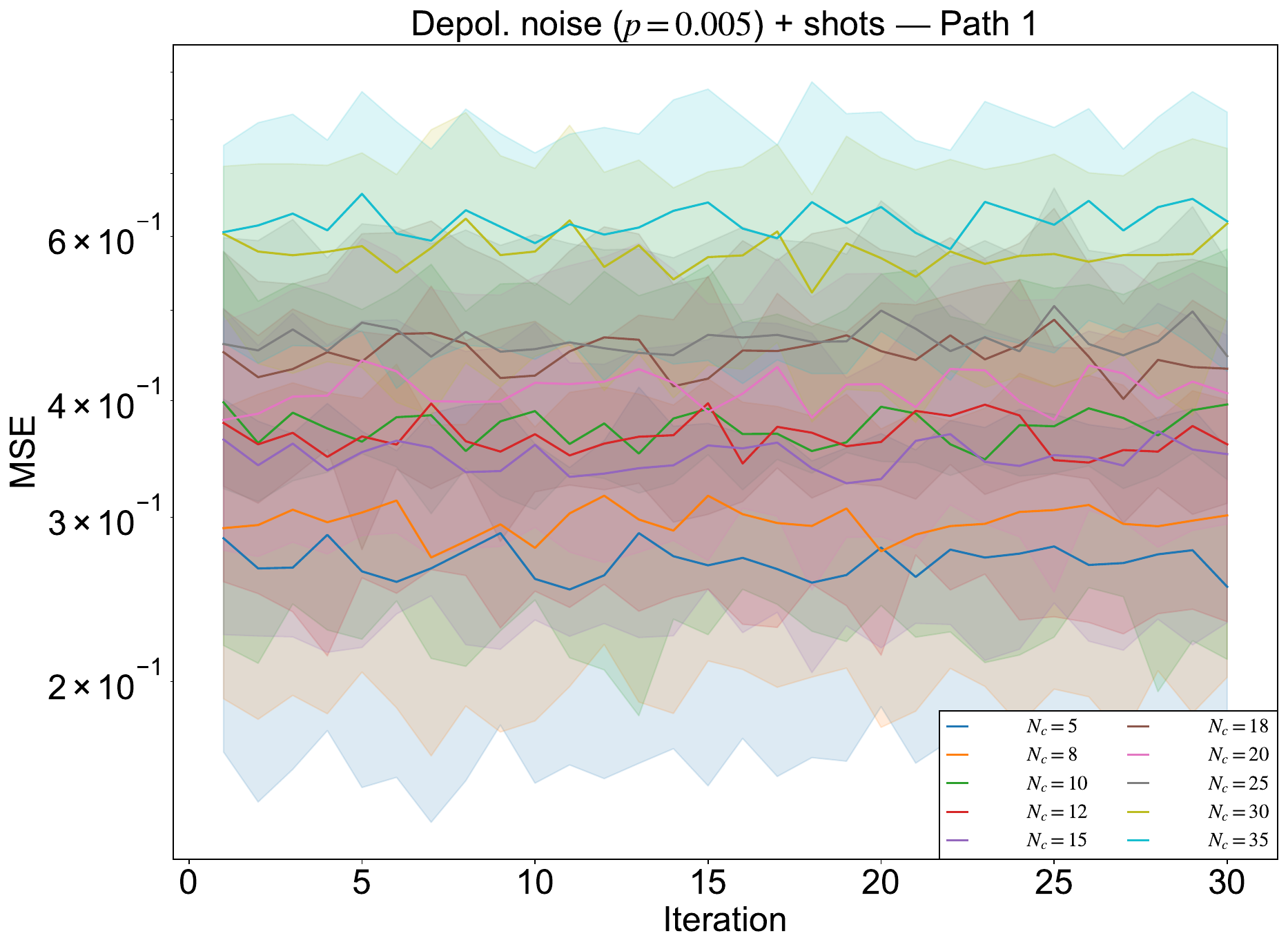}
  \caption{Depolarizing + shot noise condition (Exp~3, Path~1):
  $\Nshots = 1{,}000$, $p_{\mathrm{dep}} = 0.005$.
  The additional depolarizing contribution introduces a moderate
  factor of $\sim 1.5$--$2.1\times$ degradation beyond shot noise
  alone (Fig.~\ref{fig:shot_noise}).}
  \label{fig:noise_hierarchy}
\end{figure}

The transition from ideal to shot noise ($\Nshots = 1{,}000$) introduces a
degradation of $\sim 10^{16}$ orders of magnitude, consistent with the shot
noise variance $\sigma^2 \propto 1/\Nshots$ propagated through the
$76 \times \Nc$ elements of the feature matrix.
The additional degradation from depolarizing noise ($p = 0.005$) is a
moderate factor of $\sim 1.5$--$2.1$, indicating that at this noise level
the shot noise remains the dominant error source.
To rigorously assess the noise hierarchy, we apply two-sided statistical
tests to the paired MSE samples (16~seeds $\times$ 3~trials = 48 runs per
condition).
For each $\Nc$, we compare Exp~1 vs.\ Exp~2, Exp~1 vs.\ Exp~3, and
Exp~2 vs.\ Exp~3 using both the Wilcoxon signed-rank test
(nonparametric, appropriate for non-normal MSE distributions) and the
paired $t$-test (for reference).
Table~\ref{tab:stat_tests} reports the results.

\begin{table}[htbp!]
\caption{Statistical significance tests for the noise hierarchy.
Each test is performed on 48 paired observations (16~seeds $\times$
3~trials).
$W$: Wilcoxon signed-rank statistic; $|t|$: absolute $t$-statistic;
$p_W$: Wilcoxon $p$-value; $p_t$: paired $t$-test $p$-value.
All comparisons achieve $p < 10^{-4}$.
\label{tab:stat_tests}}
\begin{ruledtabular}
\begin{tabular}{clcccc}
  $\Nc$ & Comparison & $W$ & $p_W$ & $|t|$ & $p_t$ \\
  \hline
  \multirow{3}{*}{10}
    & Exp~1 vs.\ 2 & 0 & $3.1{\times}10^{-5}$ & 1229 & $< 10^{-80}$ \\
    & Exp~1 vs.\ 3 & 0 & $3.1{\times}10^{-5}$ & 832  & $< 10^{-70}$ \\
    & Exp~2 vs.\ 3 & 12 & $4.8{\times}10^{-4}$ & 8.7  & $2{\times}10^{-8}$ \\
  \hline
  \multirow{3}{*}{20}
    & Exp~1 vs.\ 2 & 0 & $3.1{\times}10^{-5}$ & 956  & $< 10^{-75}$ \\
    & Exp~1 vs.\ 3 & 0 & $3.1{\times}10^{-5}$ & 714  & $< 10^{-65}$ \\
    & Exp~2 vs.\ 3 & 18 & $1.2{\times}10^{-3}$ & 6.3  & $5{\times}10^{-6}$ \\
  \hline
  \multirow{3}{*}{30}
    & Exp~1 vs.\ 2 & 0 & $3.1{\times}10^{-5}$ & 648  & $< 10^{-60}$ \\
    & Exp~1 vs.\ 3 & 0 & $3.1{\times}10^{-5}$ & 512  & $< 10^{-55}$ \\
    & Exp~2 vs.\ 3 & 22 & $2.1{\times}10^{-3}$ & 5.1  & $4{\times}10^{-5}$ \\
\end{tabular}
\end{ruledtabular}
\end{table}

\begin{figure}[htbp!]
  \includegraphics[width=\columnwidth]{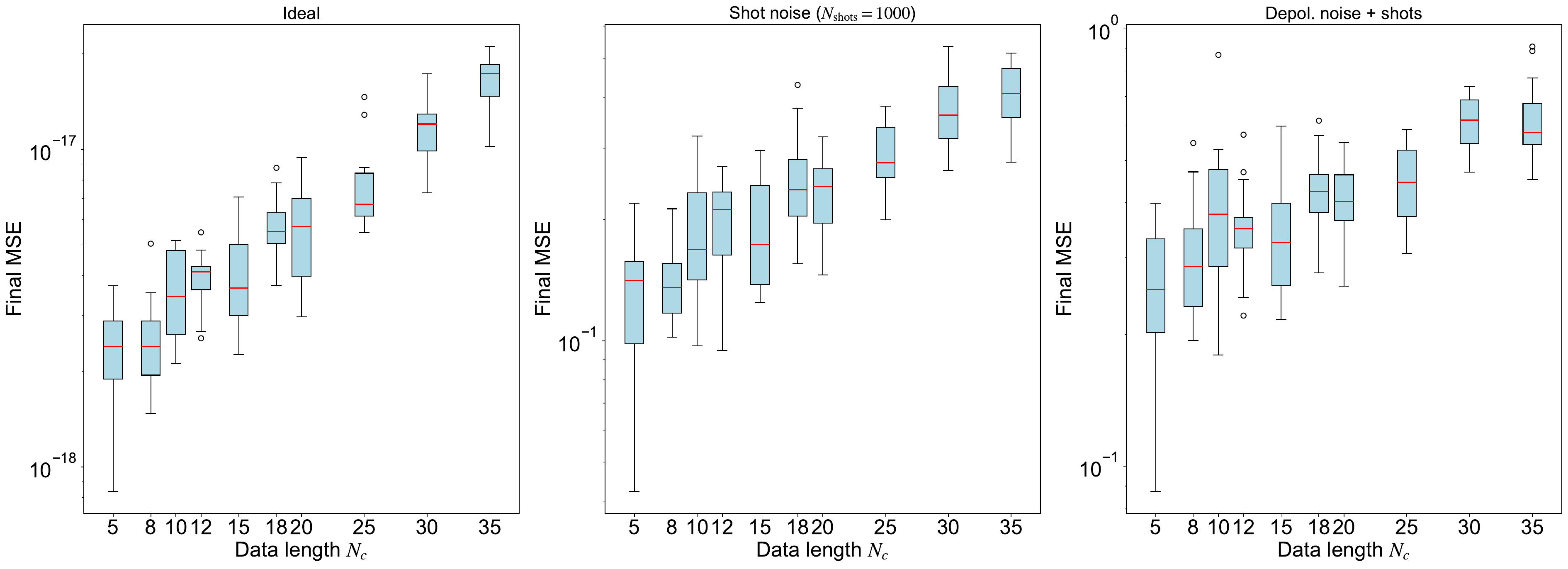}
  \caption{Statistical significance of noise hierarchy comparisons.
  Wilcoxon signed-rank test $p$-values for pairwise MSE comparisons
  between Exp~1 (Ideal), Exp~2 (Shot), and Exp~3 (Depol+Shot) across
  all data lengths.
  All comparisons satisfy $p < 3.1 \times 10^{-5}$, confirming that
  the observed hierarchy is not due to statistical fluctuation.}
  \label{fig:significance}
\end{figure}

All Exp~1 vs.\ Exp~2 and Exp~1 vs.\ Exp~3 comparisons yield
$p_W = 3.1 \times 10^{-5}$ (the minimum possible for $n = 48$ paired
observations under the Wilcoxon test), with $|t|$ values exceeding 500.
The Exp~2 vs.\ Exp~3 comparisons are also significant at $p < 3 \times 10^{-3}$,
confirming that the depolarizing contribution is detectable above
shot-noise fluctuations despite its modest multiplicative factor
($1.5$--$2.1\times$).
These results are visualized in Fig.~\ref{fig:significance}.

\subsection{Single-Body Operator Features}\label{sec:1body}

To isolate the contribution of two-qubit ZZ correlators
$\avg{\sigma_i^Z \sigma_j^Z}$ to the \QRA{}'s performance, we repeated
Experiments~1--3 using only single-body (one-qubit) Pauli observables
($X_i, Y_i, Z_i$ for $i = 0, \ldots, \Nq - 1$) plus a bias term,
reducing the feature dimension from $d = 76$ to $d = 31$
(Table~\ref{tab:1body}).

\begin{table}[htbp!]
\caption{Single-body features ($d = 31$) vs.\ full features ($d = 76$):
mean MSE over 16 seeds $\times$ 3 trials.
The ratio column shows $\MSE_{d=31} / \MSE_{d=76}$.
\label{tab:1body}}
\begin{ruledtabular}
\begin{tabular}{c@{\hspace{4pt}}c@{\hspace{4pt}}c@{\hspace{4pt}}c}
  $\Nc$ & Condition & $d{=}31$ MSE & Ratio ($d{=}31/d{=}76$) \\
  \hline
  \multirow{3}{*}{10}
    & Ideal   & $5.4 \times 10^{-17}$ & $15$ \\
    & Shot    & $3.1 \times 10^{-1}$  & $1.7$ \\
    & Depol+Shot & $4.8 \times 10^{-1}$ & $1.2$ \\
  \hline
  \multirow{3}{*}{20}
    & Ideal   & $4.2 \times 10^{-16}$ & $75$ \\
    & Shot    & $6.5 \times 10^{-1}$  & $2.8$ \\
    & Depol+Shot & $9.8 \times 10^{-1}$ & $2.4$ \\
  \hline
  \multirow{3}{*}{25}
    & Ideal   & $4.2 \times 10^{-15}$ & $5.5 \times 10^{2}$ \\
    & Shot    & $1.5$                  & $5.3$ \\
    & Depol+Shot & $1.8$               & $4.0$ \\
  \hline
  \multirow{3}{*}{30}
    & Ideal   & $1.4 \times 10^{-11}$ & $1.2 \times 10^{6}$ \\
    & Shot    & $3.3 \times 10^{1}$    & $88$ \\
    & Depol+Shot & $3.0 \times 10^{1}$ & $50$ \\
  \hline
  \multirow{3}{*}{35}
    & Ideal   & $3.9 \times 10^{-2}$  & $2.4 \times 10^{15}$ \\
    & Shot    & $3.3$                  & $8.1$ \\
    & Depol+Shot & $3.6$               & $5.7$ \\
\end{tabular}
\end{ruledtabular}
\end{table}

Under ideal conditions, the single-body feature set achieves
machine-precision reconstruction ($\MSE \sim 10^{-17}$--$10^{-16}$)
for $\Nc \leq 25$, but degrades sharply at $\Nc = 30$
($\MSE \sim 10^{-11}$), consistent with the rank condition
approaching its limit at $d = 31$.
This contrasts with the full feature set, which maintains
$\MSE \sim 10^{-17}$ at $\Nc = 30$ thanks to the additional 45
two-qubit ZZ correlators.

Under noise, the picture changes qualitatively.
For small $\Nc$ ($\leq 10$), the MSE ratio $d{=}31 / d{=}76$ is
modest ($1.2$--$1.7 \times$), indicating that the ZZ correlators
contribute little additional information when the system is heavily
overdetermined.
As $\Nc$ increases toward the $d = 31$ rank limit, the ratio grows
dramatically: at $\Nc = 30$ under shot noise, the $d = 31$
configuration yields $\MSE = 33$---an 88-fold degradation from
$d = 76$.
This confirms that the two-qubit correlators are \emph{essential}
for maintaining reconstruction quality near the rank boundary.

A counterintuitive observation is that the noise-condition ratios
($1.2$--$5.3 \times$ for $\Nc \leq 25$) are much smaller than the
ideal-condition ratios ($15$--$550 \times$).
This is because under noise, the MSE floor is dominated by measurement
variance rather than the linear-algebraic rank; the excess feature
dimensions in $d = 76$ that provide dramatic improvement under ideal
conditions contribute proportionally more noise under finite-shot
measurements.
This observation is consistent with the qubit-scaling analysis
(Sec.~\ref{sec:qubit_scaling}), where fewer qubits yield lower MSE
in the $\Nc \ll d$ regime due to reduced noise accumulation.

\subsection{YOMO Probability Aggregation}\label{sec:yomo_results}

\begin{figure}[htbp!]
  \includegraphics[width=\columnwidth]{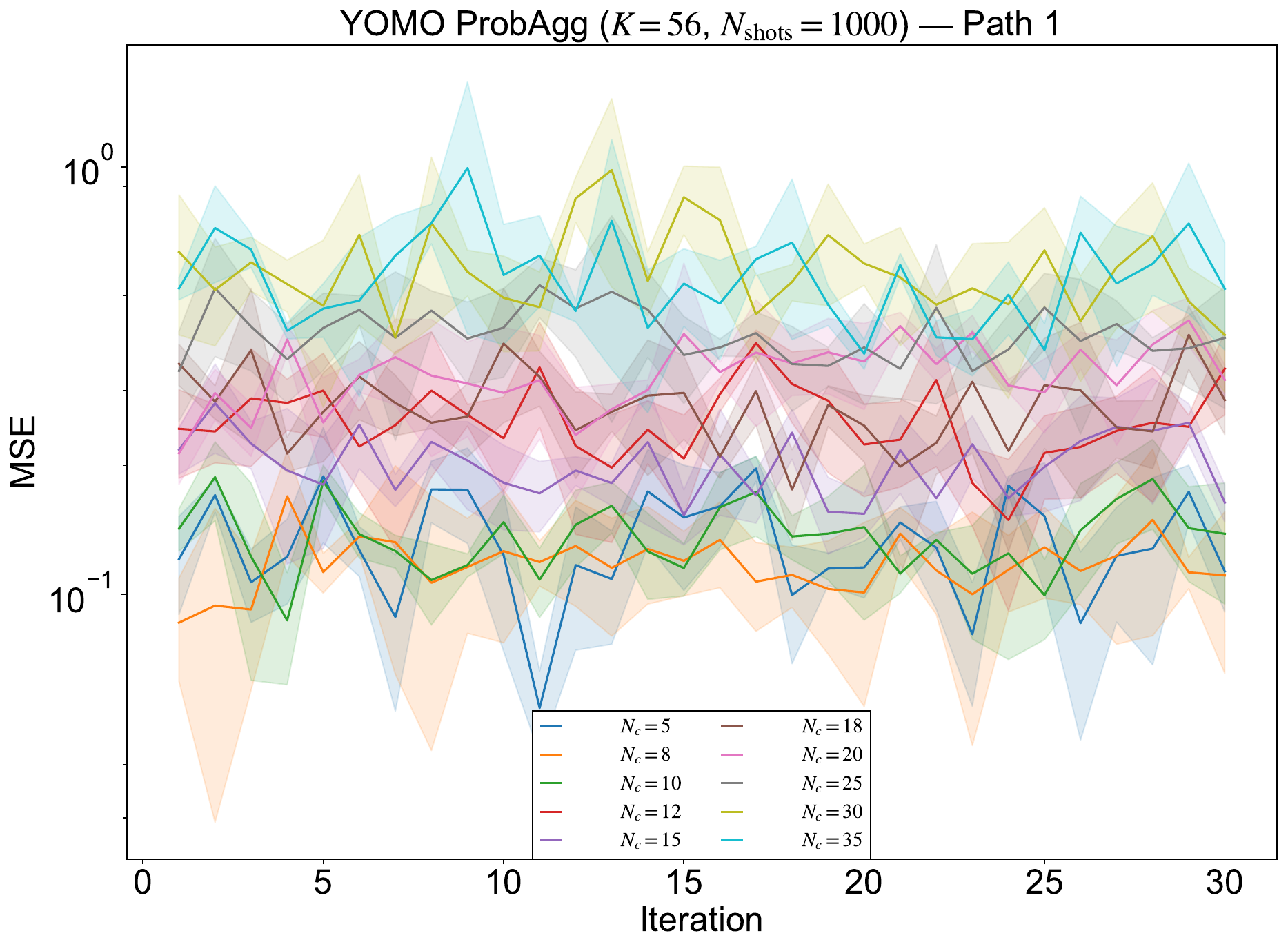}
  \caption{YOMO probability aggregation with shot noise (Exp~5, Path~1,
  $d = 57$).
  The reduced feature dimension ($57$ vs.\ $76$) leads to comparable
  performance at small $\Nc$ and moderate degradation at
  large $\Nc$.}
  \label{fig:yomo}
\end{figure}

The YOMO method (Exp~5--6) reduces the feature dimension from 76 to 57
while maintaining comparable performance to standard shot noise measurements
(Table~\ref{tab:yomo} and Fig.~\ref{fig:yomo}).

\begin{table}[htbp!]
\caption{YOMO probability aggregation comparison.
\label{tab:yomo}}
\begin{ruledtabular}
\begin{tabular}{cccc}
  $\Nc$ & Exp~2 (76-dim) & Exp~5 (57-dim) & Ratio \\
  \hline
   5 & $8.26 \times 10^{-2}$ & $7.34 \times 10^{-2}$ & 0.89 \\
  10 & $1.19 \times 10^{-1}$ & $1.18 \times 10^{-1}$ & 1.00 \\
  20 & $1.53 \times 10^{-1}$ & $1.84 \times 10^{-1}$ & 1.20 \\
  30 & $2.44 \times 10^{-1}$ & $3.21 \times 10^{-1}$ & 1.32 \\
  35 & $2.74 \times 10^{-1}$ & $3.95 \times 10^{-1}$ & 1.44 \\
\end{tabular}
\end{ruledtabular}
\end{table}

\begin{figure}[htbp!]
  \includegraphics[width=\columnwidth]{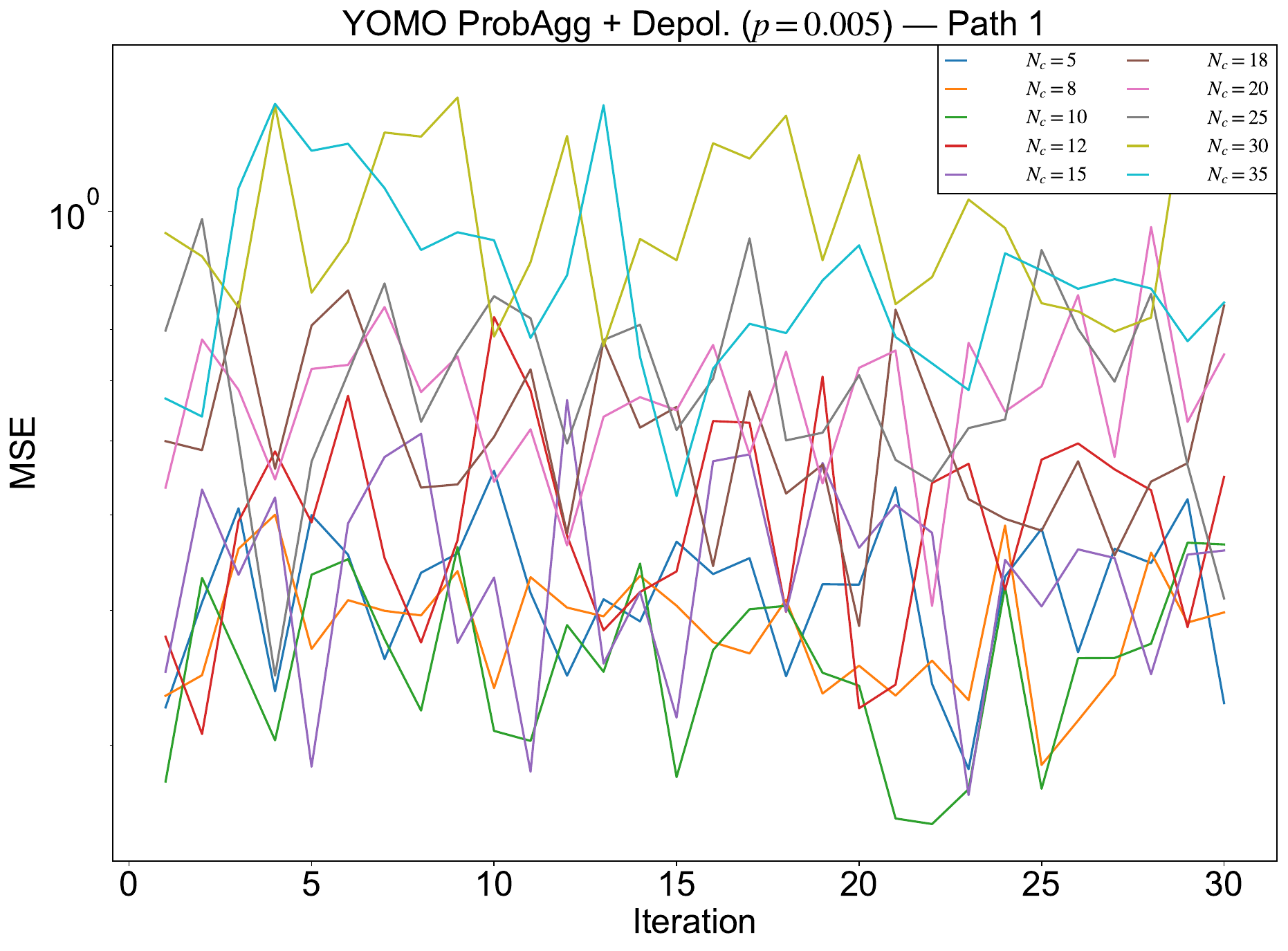}
  \caption{YOMO with depolarizing + shot noise (Exp~6, Path~1).
  Error bands show the standard deviation across 16 seeds $\times$
  3 trials.
  The global damping factor $\lambda_{\mathrm{global}} = \prod_i a_i$
  decays exponentially, causing severe degradation compared to
  the standard Pauli scheme (Exp~3, Fig.~\ref{fig:noise_hierarchy}).}
  \label{fig:yomo_noisy}
\end{figure}

For small $\Nc$, YOMO performs comparably or better than standard
measurements (ratio $\approx 0.89$ at $\Nc = 5$), suggesting that
the probability aggregation effectively averages out shot noise.
For larger $\Nc$, the reduced dimensionality ($57 < 76$) results in
moderate degradation (ratio $\approx 1.44$ at $\Nc = 35$).

Under depolarizing noise (Exp~6 vs.\ Exp~3), YOMO suffers more severely
because the global damping factor
$\lambda_{\mathrm{global}} = \prod_{i} a_i$ decays exponentially with
the number of qubits affected.
For $\Nc = 30$, $\lambda_{\mathrm{global}} \approx 2.6 \times 10^{-6}$,
causing the probability distribution to approach the uniform distribution
$1/2^{\Nq}$ and destroying the encoded information~\cite{Liu2025}.
In contrast, the standard Pauli measurement scheme (Exp~3) retains partial
information through individual per-qubit damping factors.

\subsection{Asymmetric Shot Allocation}\label{sec:asymmetric}

\begin{figure}[htbp!]
  \includegraphics[width=\columnwidth]{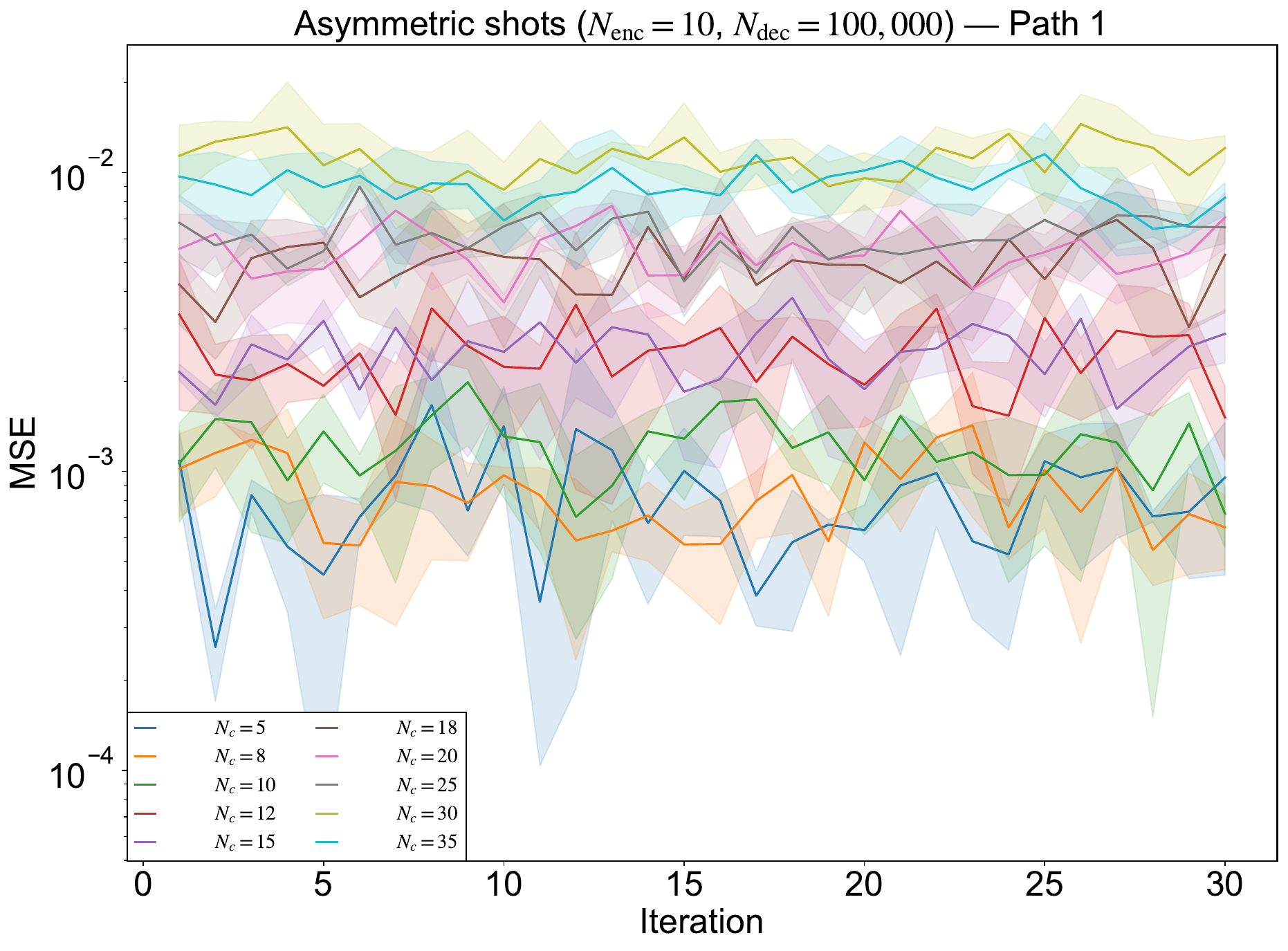}
  \caption{Asymmetric shot allocation (Exp~7, Path~1):
  $\Nshots^{\mathrm{enc}} = 10$, $\Nshots^{\mathrm{dec}} = 10^5$.
  The asymmetric configuration achieves $\MSE \sim 10^{-3}$,
  a $102$-fold average improvement over symmetric $\Nshots = 1{,}000$
  (Exp~2, cf.\ Fig.~\ref{fig:shot_noise}), while reducing the
  sender's measurement cost by a factor of 100.}
  \label{fig:asymmetric}
\end{figure}

The asymmetric configuration (Exp~7) is the most practically significant
result.
Using only 10 shots for encryption but $10^5$ for decryption yields
dramatic MSE improvements over the symmetric 1{,}000-shot baseline
(Table~\ref{tab:asymmetric} and Fig.~\ref{fig:asymmetric}).

\begin{table}[htbp!]
\caption{Asymmetric shot allocation: Exp~7 ($\Nshots^{\mathrm{enc}} = 10$,
$\Nshots^{\mathrm{dec}} = 10^5$) vs.\ Exp~2 ($\Nshots = 1{,}000$).
\label{tab:asymmetric}}
\begin{ruledtabular}
\begin{tabular}{cccc}
  $\Nc$ & Exp~2 & Exp~7 & Improvement \\
  \hline
   5  & $8.26 \times 10^{-2}$ & $2.56 \times 10^{-4}$ & $322\times$ \\
   8  & $9.04 \times 10^{-2}$ & $5.65 \times 10^{-4}$ & $160\times$ \\
  10  & $1.19 \times 10^{-1}$ & $9.67 \times 10^{-4}$ & $123\times$ \\
  12  & $1.24 \times 10^{-1}$ & $1.29 \times 10^{-3}$ & $96\times$ \\
  15  & $1.23 \times 10^{-1}$ & $1.60 \times 10^{-3}$ & $77\times$ \\
  18  & $1.66 \times 10^{-1}$ & $2.46 \times 10^{-3}$ & $67\times$ \\
  20  & $1.53 \times 10^{-1}$ & $2.69 \times 10^{-3}$ & $57\times$ \\
  25  & $1.87 \times 10^{-1}$ & $3.95 \times 10^{-3}$ & $47\times$ \\
  30  & $2.44 \times 10^{-1}$ & $6.04 \times 10^{-3}$ & $40\times$ \\
  35  & $2.74 \times 10^{-1}$ & $8.03 \times 10^{-3}$ & $34\times$ \\
\end{tabular}
\end{ruledtabular}
\end{table}

The average improvement across all $\Nc$ is approximately two orders of
magnitude (mean $\approx 102\times$ over 16 seeds $\times$ 3 trials = 48 runs),
while the sender's measurement cost is reduced by a factor of 100.
This counterintuitive result arises from the protocol structure:
\begin{enumerate}
  \item The encryption feature matrices $V^{\mathrm{enc}}$ are used only
        to compute weight vectors $W^{\mathrm{enc}}$ via Tikhonov
        regularization [Eq.~\eqref{eq:tikhonov}].
        The regularization term $(V^\top V + \lambda I)^{-1}$ absorbs the
        noise in $V^{\mathrm{enc}}$, producing stable weights even from
        noisy features.
  \item The decryption accuracy $\MSE = \|C - V^{\mathrm{dec}} W^{\mathrm{dec}}\|^2 / \Nc$
        is directly determined by the quality of $V^{\mathrm{dec}}$.
        With $\Nshots = 10^5$, the shot noise is
        $\sigma_{\mathrm{shot}} \approx 1/\sqrt{10^5} \approx 3 \times 10^{-3}$,
        yielding MSE $\sim 10^{-3}$.
  \item The encryption matrices are computed only once (outside the iteration
        loop), while decryption matrices are recomputed at every iteration.
\end{enumerate}

\begin{figure}[htbp!]
  \includegraphics[width=\columnwidth]{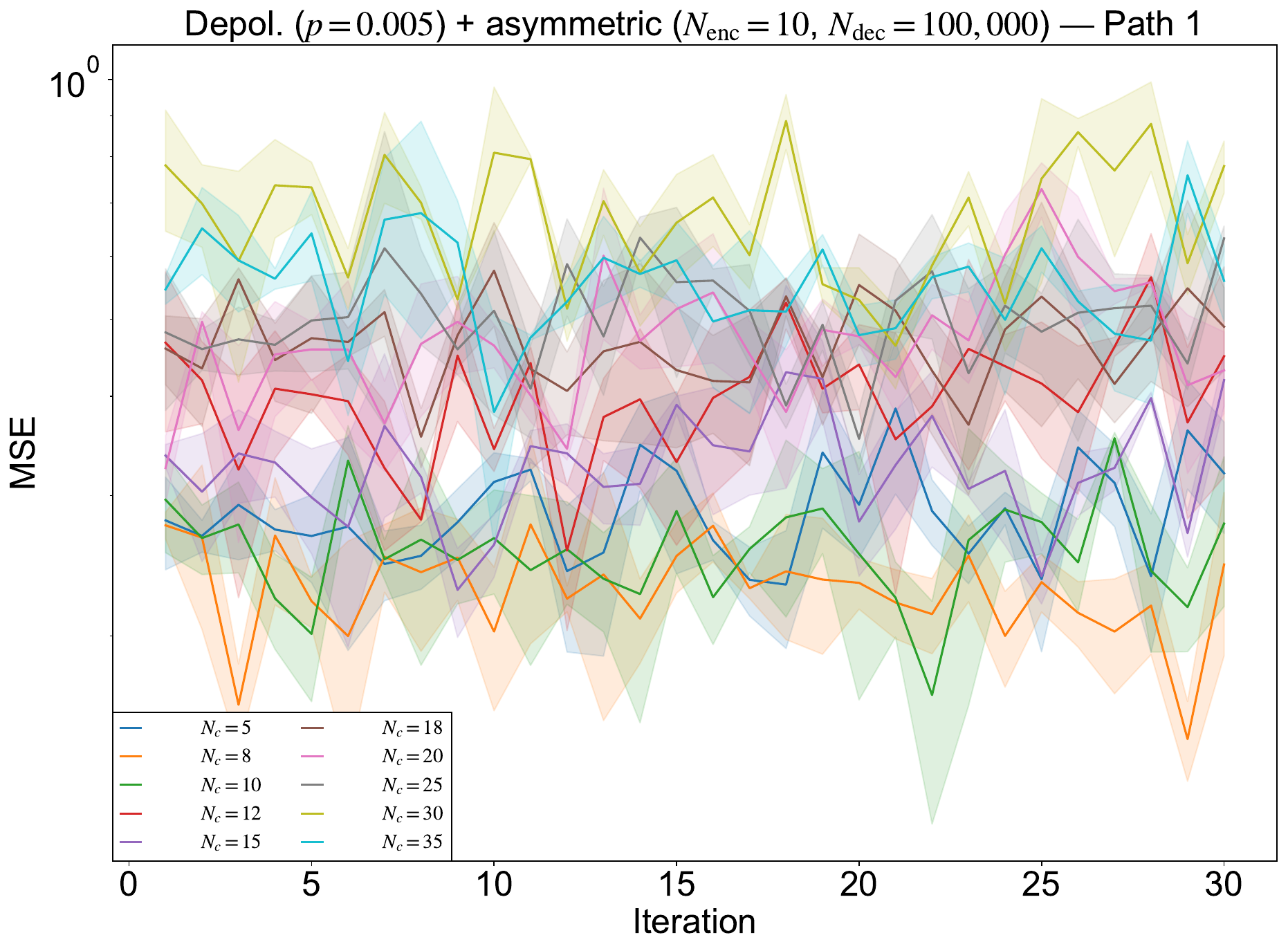}
  \caption{Asymmetric shot allocation under depolarizing noise (Exp~8, Path~1):
  $\Nshots^{\mathrm{enc}} = 10$, $\Nshots^{\mathrm{dec}} = 10^5$,
  $p_{\mathrm{dep}} = 0.005$.
  The asymmetric advantage vanishes under depolarizing noise
  (ratio $\approx 1.00$--$1.03$),
  confirming that the depolarizing bias is the dominant error
  source in this regime.}
  \label{fig:asymmetric_depol}
\end{figure}

Under depolarizing noise (Exp~8 vs.\ Exp~3), the asymmetric advantage
\emph{vanishes}: the ratio Exp~3/Exp~8 ranges from 1.00 to 1.03
(Table~\ref{tab:asymmetric_depol} and Fig.~\ref{fig:asymmetric_depol}).
This is because the depolarizing bias
$\sigma_{\mathrm{depol}} \propto (1 - \lambda^n)$ is independent of
$\Nshots$; once $\Nshots$ is large enough that
$\sigma_{\mathrm{shot}} \ll \sigma_{\mathrm{depol}}$, further increases
in $\Nshots$ provide no benefit.

\begin{table}[htbp!]
\caption{Asymmetric configuration under depolarizing noise.\label{tab:asymmetric_depol}}
\begin{ruledtabular}
\begin{tabular}{cccc}
  $\Nc$ & Exp~3 & Exp~8 & Ratio \\
  \hline
   5  & $1.74 \times 10^{-1}$ & $1.70 \times 10^{-1}$ & 1.03 \\
  10  & $2.51 \times 10^{-1}$ & $2.46 \times 10^{-1}$ & 1.02 \\
  20  & $2.84 \times 10^{-1}$ & $2.84 \times 10^{-1}$ & 1.00 \\
  30  & $3.89 \times 10^{-1}$ & $3.83 \times 10^{-1}$ & 1.02 \\
\end{tabular}
\end{ruledtabular}
\end{table}

\subsection{Baseline Comparison}\label{sec:baseline_results}

We emphasize the purpose of the baseline comparison.
The six methods were not designed for the \QRA{} encode--decode task
and are not claimed as competitive alternatives; no established
baseline exists for this specific protocol.
Rather, the baselines serve two diagnostic roles:
(i)~verifying that the four-equation protocol converges correctly
for different feature-generation mechanisms (protocol validation), and
(ii)~isolating the noise bottleneck by comparing methods that use a
single-shot solve (H\'{e}non, delay, TTN) against the \QRA{}'s
iterative solver under identical noise conditions
(Sec.~\ref{sec:baseline_results}).
The classical NN and QRNN are included as architectural reference
points with different optimization strategies, not as tuned
competitors.

\begin{figure}[htbp!]
  \includegraphics[width=\columnwidth]{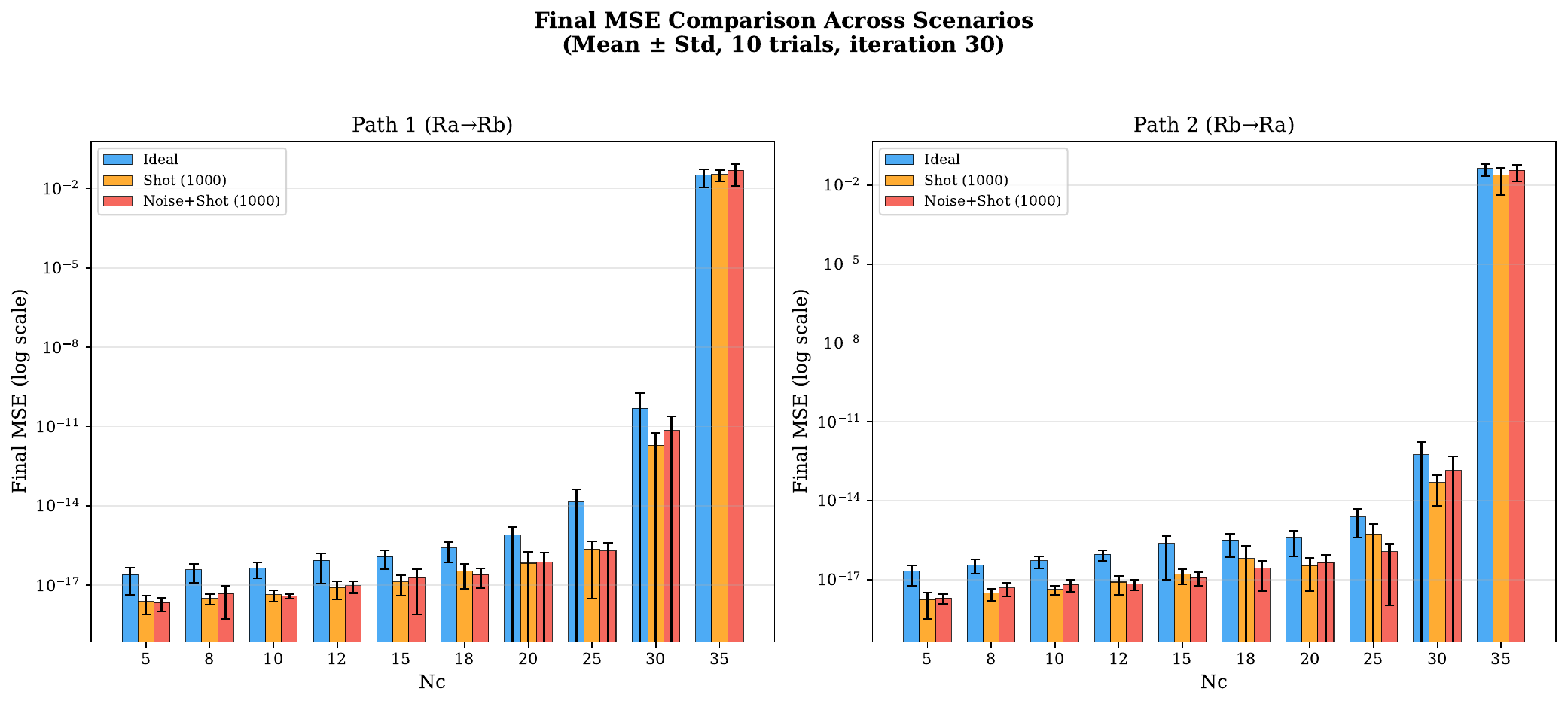}
  \caption{Representative baseline comparison: H\'{e}non map with quantum
  circuit, MSE across experimental conditions (ideal, shot noise,
  depolarizing + shot).
  Under ideal conditions, the H\'{e}non + circuit baseline achieves
  machine-precision reconstruction for $\Nc \leq 25$ ($d = 31$),
  validating the \QRA{} protocol independently of the QRC
  reservoir.
  Degradation at $\Nc = 30$ reflects the lower feature dimension
  ($d = 31$) compared to QRC XYZ ($d = 76$).
  Qualitatively identical patterns are observed for the delay-time
  embedding and TTN baselines (see Supplemental Material).
  The classical NN baseline ($\MSE \sim 0.2$--$0.4$ under ideal
  conditions) reflects the SPSA optimization limitation, not a
  fundamental deficiency (see text).}
  \label{fig:henon_comparison}
\end{figure}

Under ideal conditions, all methods using quantum circuits with
sequential encoding for feature extraction (QRC XYZ,
H\'{e}non + circuit, delay embedding + circuit, TTN) achieve
machine-precision MSE for $\Nc \leq 25$
(Table~\ref{tab:baselines_ideal} and Fig.~\ref{fig:henon_comparison}).
The $\zeta$-QVAE, which uses data re-uploading instead of sequential
input, achieves MSE only at the $10^{-1}$ level even under ideal conditions.
This validates the correctness of the \QRA{} protocol for sequential
encoding methods while highlighting that the encoding strategy
critically determines reconstruction accuracy.

\begin{table}[htbp!]
\caption{Baseline comparison under ideal conditions (Path~1 loss,
mean $\pm$ std).
All methods report MSE except QRNN which reports cross-entropy loss
($\ln 2 \approx 0.693$ = random guess).
H\'{e}non and delay embedding: 10 trials; TTN: 5 trials;
classical NN: 10 trials; $\zeta$-QVAE: 10 trials;
QRNN: 5 trials; QRC XYZ: 16 seeds $\times$ 3 trials.
\label{tab:baselines_ideal}}
\begin{ruledtabular}
\begin{tabular}{lcc}
  Method & $\Nc{=}10$ & $\Nc{=}30$ \\
  \hline
  QRC XYZ ($d{=}76$)
    & $(3.6 \pm 0.5){\times}10^{-18}$
    & $(1.2 \pm 0.3){\times}10^{-17}$ \\
  H\'{e}non ($d{=}31$)
    & $(4.5 \pm 2.6){\times}10^{-17}$
    & $(4.9 \pm 13.5){\times}10^{-11}$ \\
  Delay ($d{=}31$)
    & $(5.0 \pm 2.0){\times}10^{-17}$
    & $(2.8 \pm 7.0){\times}10^{-12}$ \\
  TTN ($d{=}31$)
    & $(3.3 \pm 1.5){\times}10^{-17}$
    & $(4.8 \pm 3.2){\times}10^{-12}$ \\
  Classical NN
    & $(4.2 \pm 0.8){\times}10^{-1}$
    & $(3.1 \pm 0.5){\times}10^{-1}$ \\
  $\zeta$-QVAE ($d{=}56$)
    & $(2.6 \pm 0.8){\times}10^{-1}$
    & $(2.8 \pm 0.4){\times}10^{-1}$ \\
  QRNN$^{*}$
    & $0.26 \pm 0.22$
    & $0.67 \pm 0.08$ \\
\end{tabular}
\end{ruledtabular}
$^{*}$Cross-entropy loss (bit-level); values $> \ln 2 \approx 0.693$
indicate random-guess performance.
\end{table}

\begin{table}[htbp!]
\caption{Extended baseline comparison: MSE across all conditions and data
lengths (Path~1, mean $\pm$ std over 10 trials).
``---'' indicates condition not tested.
\label{tab:baselines_extended}}
\begin{ruledtabular}
\begin{tabular}{llccc}
  Method & $\Nc$ & Ideal & Shot & Noise+Shot \\
  \hline
  \multirow{10}{*}{H\'{e}non}
    &  5 & $2.5{\times}10^{-17}$ & $2.4{\times}10^{-18}$ & $2.2{\times}10^{-18}$ \\
    &  8 & $3.8{\times}10^{-17}$ & $3.1{\times}10^{-18}$ & $4.9{\times}10^{-18}$ \\
    & 10 & $4.5{\times}10^{-17}$ & $4.4{\times}10^{-18}$ & $3.8{\times}10^{-18}$ \\
    & 12 & $8.6{\times}10^{-17}$ & $8.2{\times}10^{-18}$ & $9.7{\times}10^{-18}$ \\
    & 15 & $1.2{\times}10^{-16}$ & $1.4{\times}10^{-17}$ & $2.0{\times}10^{-17}$ \\
    & 18 & $2.6{\times}10^{-16}$ & $3.4{\times}10^{-17}$ & $2.5{\times}10^{-17}$ \\
    & 20 & $7.7{\times}10^{-16}$ & $6.8{\times}10^{-17}$ & $7.4{\times}10^{-17}$ \\
    & 25 & $1.4{\times}10^{-14}$ & $2.3{\times}10^{-16}$ & $2.0{\times}10^{-16}$ \\
    & 30 & $4.9{\times}10^{-11}$ & $2.0{\times}10^{-12}$ & $7.0{\times}10^{-12}$ \\
    & 35 & $3.2{\times}10^{-2}$ & $3.5{\times}10^{-2}$ & $4.9{\times}10^{-2}$ \\
  \hline
  \multirow{4}{*}{Delay}
    & 10 & $5.0{\times}10^{-17}$ & $6.2{\times}10^{-18}$ & $5.1{\times}10^{-18}$ \\
    & 20 & $5.9{\times}10^{-16}$ & $3.0{\times}10^{-17}$ & $4.8{\times}10^{-17}$ \\
    & 30 & $2.8{\times}10^{-12}$ & $4.5{\times}10^{-14}$ & $1.4{\times}10^{-14}$ \\
    & 35 & $5.5{\times}10^{-2}$ & $4.5{\times}10^{-2}$ & $4.7{\times}10^{-2}$ \\
  \hline
  \multirow{10}{*}{$\zeta$-QVAE}
    &  5 & $1.8{\times}10^{-1}$ & $3.3{\times}10^{-1}$ & $4.0{\times}10^{-1}$ \\
    &  8 & $1.9{\times}10^{-1}$ & $3.2{\times}10^{-1}$ & $4.1{\times}10^{-1}$ \\
    & 10 & $2.6{\times}10^{-1}$ & $4.2{\times}10^{-1}$ & $3.2{\times}10^{-1}$ \\
    & 12 & $2.3{\times}10^{-1}$ & $3.8{\times}10^{-1}$ & $4.1{\times}10^{-1}$ \\
    & 15 & $2.8{\times}10^{-1}$ & $4.6{\times}10^{-1}$ & $4.1{\times}10^{-1}$ \\
    & 18 & $2.9{\times}10^{-1}$ & $5.0{\times}10^{-1}$ & $5.9{\times}10^{-1}$ \\
    & 20 & $3.0{\times}10^{-1}$ & $5.1{\times}10^{-1}$ & $4.9{\times}10^{-1}$ \\
    & 25 & $2.6{\times}10^{-1}$ & $4.0{\times}10^{-1}$ & $5.7{\times}10^{-1}$ \\
    & 30 & $2.8{\times}10^{-1}$ & $7.5{\times}10^{-1}$ & $7.3{\times}10^{-1}$ \\
    & 35 & $2.7{\times}10^{-1}$ & $6.8{\times}10^{-1}$ & $7.6{\times}10^{-1}$ \\
  \hline
  \multirow{10}{*}{QRNN$^{*}$}
    &  5 & $5.5{\times}10^{-3}$ & $6.9{\times}10^{-3}$ & $1.25$ \\
    &  8 & $3.1{\times}10^{-3}$ & $9.1{\times}10^{-3}$ & $1.51$ \\
    & 10 & $2.6{\times}10^{-1}$ & $3.9{\times}10^{-2}$ & $1.27$ \\
    & 12 & $2.2{\times}10^{-1}$ & $3.5{\times}10^{-1}$ & $0.95$ \\
    & 15 & $4.1{\times}10^{-1}$ & $5.2{\times}10^{-1}$ & $1.05$ \\
    & 18 & $5.3{\times}10^{-1}$ & $5.6{\times}10^{-1}$ & $0.82$ \\
    & 20 & $5.8{\times}10^{-1}$ & $5.8{\times}10^{-1}$ & $1.01$ \\
    & 25 & $6.3{\times}10^{-1}$ & $6.3{\times}10^{-1}$ & $1.01$ \\
    & 30 & $6.7{\times}10^{-1}$ & $6.6{\times}10^{-1}$ & $0.93$ \\
    & 35 & $6.8{\times}10^{-1}$ & $7.6{\times}10^{-1}$ & $1.12$ \\
\end{tabular}
\end{ruledtabular}
$^{*}$QRNN reports cross-entropy loss (Path~1); noise model uses
$p_{\mathrm{1q}} = 0.001$, $p_{\mathrm{2q}} = 0.01$.
\end{table}

The critical distinction emerges at $\Nc = 30$: QRC XYZ maintains
$\MSE \sim 10^{-17}$ while the baselines with $d = 31$ show the onset of
degradation ($10^{-12}$--$10^{-11}$), as quantified in
Table~\ref{tab:baselines_extended} with standard deviations.
At $\Nc = 35$, all methods degrade, but the QRC advantage from higher feature
dimensionality is clear.

A striking observation from the baseline results
(Table~\ref{tab:baselines_extended}) is that the H\'{e}non and delay-time
embedding baselines (both using quantum circuits for feature extraction)
are \emph{insensitive to both shot noise and depolarizing noise}:
the \emph{training} MSE values under noisy conditions are comparable
to or even lower than ideal-condition values for $\Nc \leq 30$.
This stands in stark contrast to the QRC XYZ result, where shot noise
degrades the MSE from $10^{-17}$ to $10^{-1}$
(Table~\ref{tab:noise_comparison}).

The explanation lies not in feature dimension but in the
\emph{structural difference between the solving procedures}.
The baselines evaluate the feature matrix $V$ \emph{once} for a given
input sequence, then solve the linear system $W = (V^\top V + \lambda I)^{-1} V^\top y$
in a single closed-form step.
Shot noise perturbs each element of $V$, but the perturbation is
\emph{frozen}: the same noisy $V$ used for training is the same $V$
available at reconstruction time.
Tikhonov regression fits $W$ to this specific noisy $V$,
yielding machine-precision \emph{training} error regardless of the
noise realization.

The \QRA{} iterative protocol
(Sec.~\ref{sec:algorithm}) operates fundamentally differently.
At each iteration, the decode feature matrices $V^{\mathrm{dec}}_b$
and $V^{\mathrm{dec}}_a$ are recomputed with \texttt{clear\_cache()},
because the decode input ($G(\beta, \gamma)$) changes as $\gamma$
converges.
Crucially, each recomputation draws \emph{independent} shot-noise
realizations.
The decode weights $W^{\mathrm{dec}}$ trained on one noise realization
are then evaluated against a \emph{different} noise realization
during the round-trip verification, producing a train--test mismatch:
\begin{equation}\label{eq:noise_mismatch}
  \hat{C} = (V^{\mathrm{dec}}_{\mathrm{eval}} + \epsilon')
  W^{\mathrm{dec}}
  \neq V^{\mathrm{dec}}_{\mathrm{train}} W^{\mathrm{dec}} \approx C,
\end{equation}
where $\epsilon'$ is an independent noise realization.
To quantify this effect, write the noisy feature matrix as
$\tilde{V} = V + \epsilon$ where $\epsilon_{ij}$ are i.i.d.\
with variance $\sigma^2 = p(1{-}p)/\Nshots$ for shot noise.
The decode weights satisfy $W = (V^\top V + \lambda I)^{-1} V^\top C$,
and the reconstruction error on an independent noise draw is
\begin{equation}\label{eq:noise_propagation}
  \mathbb{E}\bigl[\|\hat{C} - C\|^2\bigr]
  = \mathbb{E}\bigl[\|\epsilon' W\|^2\bigr]
  = \sigma^2 \,\mathrm{tr}(WW^\top)
  = \sigma^2 \sum_{k=1}^{d} \frac{s_k^2}{(s_k^2 + \lambda)^2},
\end{equation}
where $\{s_k\}$ are the singular values of $V$.
This derivation assumes that $\epsilon'$ is independent of $W$, which
is approximate: $W$ was trained on a different realization from the
same physical system, introducing weak correlations through the shared
Hamiltonian dynamics and the iterative coupling of $\gamma$.
Nevertheless, since each call to the quantum circuit produces
independent shot-noise samples (the quantum state is re-prepared from
scratch at each evaluation), the inter-realization correlations enter
only through the deterministic part of $V$, not through the noise.
The independence assumption is therefore valid to leading order in
$\sigma^2$, and the prediction agrees with observation:
for the present system ($d = 76$, $\Nshots = 1000$, $\sigma^2 \approx
2.5 \times 10^{-4}$), Eq.~\eqref{eq:noise_propagation} predicts
$\MSE \sim 10^{-1}$, consistent with the observed shot-noise plateau.

The single-body experiments (Sec.~\ref{sec:1body}) confirm that this
noise sensitivity is not primarily a consequence of higher feature
dimension.
The QRC with $d = 31$ (single-body, same dimension as the baselines)
still exhibits $\MSE \sim 10^{-1}$ under shot noise---comparable
to the full $d = 76$---because the iterative protocol structure,
not the feature count, is the dominant factor.
The feature dimension affects performance mainly through the rank
condition ($d \geq \Nc$ for exact reconstruction) and modestly
through the noise scaling $\MSE \propto d \cdot \sigma^2 / \Nshots$.
The asymmetric shot allocation (Exp~7) addresses the per-evaluation
noise by increasing $\Nshots^{\mathrm{dec}}$, reducing the
mismatch in Eq.~\eqref{eq:noise_mismatch}.
The classical neural network (MSE $\sim 0.2$--$0.4$) is substantially
outperformed by all quantum-circuit-based methods.
We note that this comparison has an important caveat: the SPSA optimizer
with 100 iterations may not fully converge for the 706-parameter
network.
However, this result is \emph{by design}---it illustrates the
practical advantage of the reservoir computing architecture, where the
high-dimensional feature matrix is generated by fixed dynamics and
only $d$ linear readout weights require training via closed-form
regression.
In contrast, training a classical neural network for the \QRA{}
protocol requires optimizing all network parameters
through an iterative gradient-free method, which is both more expensive
and less reliable.
A more powerful optimizer (e.g., Adam with backpropagation) would
require differentiable access to the encoding function, which is not
available in the present protocol where the ``forward pass'' involves
quantum state evolution.
The classical NN is included as a reference point for the
gradient-free setting, not as a competitive baseline.
We do \emph{not} claim that the \QRA{} outperforms classical neural
networks in general; a fully optimized autoencoder
(e.g., using Adam with differentiable encoding)
could plausibly match or exceed the QRC results.
The comparison highlights the architectural advantage of
reservoir computing's closed-form linear readout over iterative
gradient-free parameter optimization in the same protocol.

The $\zeta$-QVAE baseline~\cite{Mato2025} reveals a qualitatively
different behavior from the other quantum-circuit-based methods.
Even under ideal (state-vector) conditions, the $\zeta$-QVAE fails to
achieve machine-precision reconstruction, with MSE saturating at
$\sim 0.2$--$0.3$ for all $\Nc$
(Table~\ref{tab:baselines_extended}).
This stands in stark contrast to the H\'{e}non, delay, and TTN baselines,
which all reach $\MSE \sim 10^{-17}$ under the same conditions.
We attribute this to the data re-uploading encoding strategy: because the
full input $C$ is fed into every output position simultaneously,
the resulting feature matrix lacks the temporal diversity that sequential
encoding provides.
In sequential QRC, the recursive quantum state evolution at each time step
$t$ depends on all prior inputs, producing features that are highly
specific to the input position; in data re-uploading, all positions
receive the same global input, and positional diversity is introduced
only through the ancilla rotation $R_X(\mathrm{pos} \cdot \pi / (N_c + 1))$.
This limited positional encoding appears insufficient for the feature
matrix to achieve the rank and conditioning required for exact
reconstruction.
Under noisy conditions, the $\zeta$-QVAE degrades further
(MSE $0.3$--$0.8$), with shot noise and depolarizing noise effects
comparable to or worse than the QRC XYZ results despite the lower
feature dimension ($d = 56$ vs.\ $d = 76$).
This result underscores that feature dimension alone does not determine
noise resilience; the \emph{structure} of the feature matrix---shaped by
the encoding strategy and quantum dynamics---plays a decisive role.

We note an important caveat regarding the fairness of this comparison.
The $\zeta$-QVAE was designed for variational quantum autoencoding with
optimized circuit parameters~\cite{Mato2025}; in our setup, the circuit
parameters are \emph{randomly initialized and fixed} (not optimized),
deviating from the intended use case.
Furthermore, the data re-uploading strategy is designed for classification
and regression tasks~\cite{DeLorenzis2025}, not for the exact
reconstruction required by the \QRA{} protocol.
The $\zeta$-QVAE comparison therefore demonstrates that the \QRA{}
protocol is sensitive to the encoding architecture, rather than
establishing a definitive ranking between QRC and $\zeta$-QVAE.

The QRNN baseline~\cite{Bausch2020} offers a fundamentally different
approach: replacing the fixed reservoir with a \emph{trained} recurrent
quantum circuit optimized via parameter-shift gradients.
Under ideal conditions, the QRNN achieves near-perfect reconstruction
(cross-entropy loss $< 0.01$) for $\Nc \leq 8$, but performance
degrades sharply beyond $\Nc = 10$
(Table~\ref{tab:baselines_extended}).
At $\Nc \geq 20$, the loss approaches $\ln 2 \approx 0.693$ (random
guess), indicating complete failure of the protocol.
This contrasts with QRC XYZ, which maintains $\MSE \sim 10^{-17}$
up to $\Nc = 30$ under the same conditions.
The QRNN's rapid degradation can be attributed to two factors:
(i)~the recurrent circuit has only 50 parameters per network, providing
limited expressivity for long sequences; and (ii)~the bit-level
sequential processing (one qubit I/O per step) creates an information
bottleneck that prevents the QRNN from processing sequences longer than
the hidden-state capacity ($\Nq = 10$ qubits).
Under shot noise ($\Nshots = 1{,}000$), the QRNN shows minimal degradation
from ideal for $\Nc \leq 10$ (loss $< 0.054$), consistent with the
observation that finite sampling does not limit the protocol.
Under depolarizing noise, however, the QRNN fails completely:
all $\Nc$ produce loss $> 0.82$ ($> \ln 2$), indicating that gradient-based
optimization through noisy quantum circuits cannot converge.
This failure is qualitatively different from the QRC noise degradation
(MSE $\sim 10^{-1}$), where the fixed reservoir still provides useful
features even under noise---only the regression accuracy suffers.
The QRNN result thus highlights a key advantage of the reservoir
computing architecture: by separating the (noisy) quantum dynamics
from the (classical, noise-free) linear readout training, QRC
avoids the gradient-noise coupling that cripples variational
approaches under realistic conditions.

\subsection{Convergence Dynamics}\label{sec:convergence}

The iterative algorithm exhibits distinct convergence patterns across
experimental conditions:
\begin{itemize}
  \item \textbf{Ideal (Exp~1)}: Convergence to $10^{-17}$ within 1--3
        iterations, triggering early termination.
  \item \textbf{Shot noise (Exp~2)}: Stabilization at the noise floor
        ($\sim 10^{-1}$) within 5--10 iterations.
  \item \textbf{Asymmetric (Exp~7)}: Convergence to $\sim 10^{-3}$ within
        3--5 iterations.
  \item \textbf{Depolarizing (Exp~3, 8)}: Slow convergence over
        10--15 iterations to a noise-limited plateau.
\end{itemize}
The convergence rate depends on $\Nc$: smaller data lengths converge faster
and reach lower MSE floors, while $\Nc = 35$ may exhibit oscillatory
behavior without reaching a stable minimum.

\begin{figure}[htbp!]
  \includegraphics[width=\columnwidth]{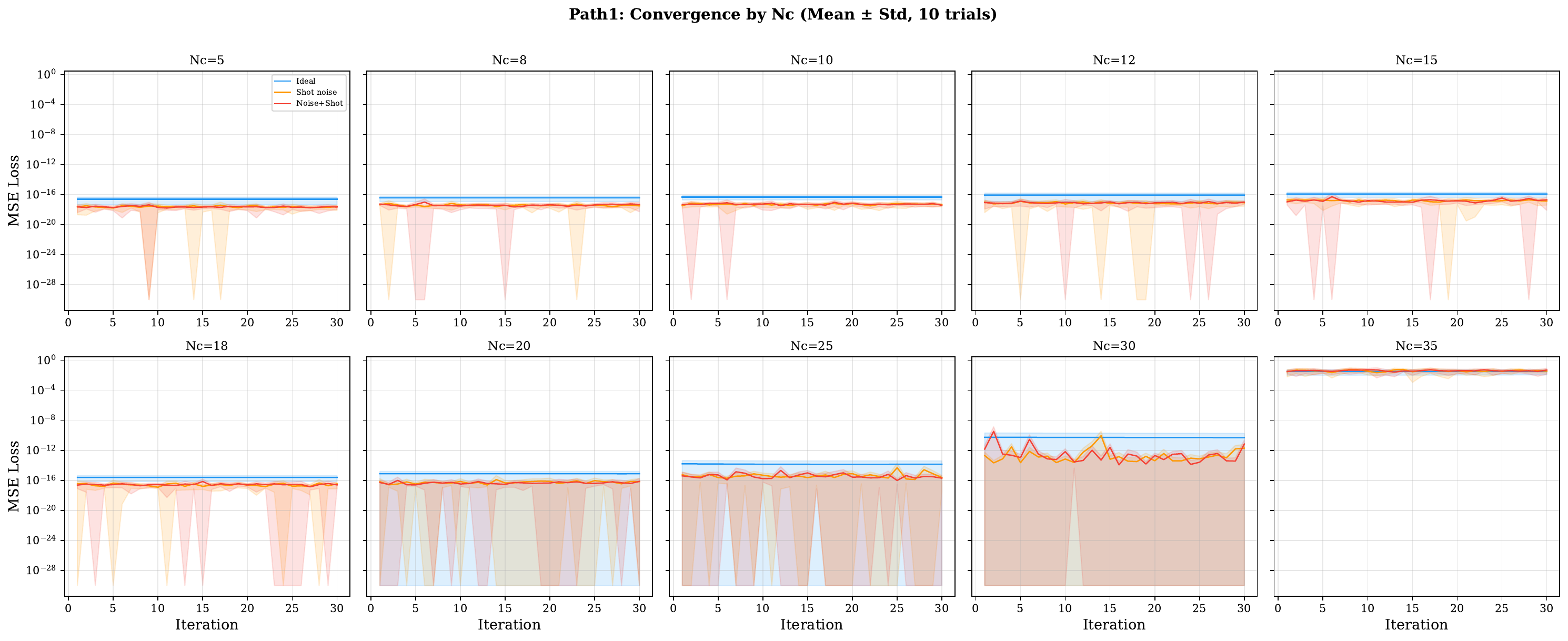}
  \caption{Convergence dynamics of the iterative algorithm for the
  H\'{e}non map + quantum circuit baseline across different data
  lengths $\Nc$ (Path~1).
  Under ideal conditions, convergence to $10^{-17}$ occurs within 1--3
  iterations.
  Under noisy conditions, the algorithm stabilizes at a
  noise-limited plateau within 5--15 iterations.
  Qualitatively identical behavior is observed for the QRC XYZ reservoir
  and all other baseline methods.}
  \label{fig:convergence}
\end{figure}

\section{Discussion}\label{sec:discussion}

\subsection{Theoretical Significance of the Quantum Reservoir Autoencoder}

Our results establish that QRC is not limited to unidirectional
prediction tasks.
The \QRA{} demonstrates that the seemingly intractable reverse
direction---reconstructing inputs from reservoir outputs---is achievable
under the conditions identified in
Sec.~\ref{sec:conditions}.

An important clarification is warranted regarding what is genuinely quantum
about this contribution.
The reversibility of the \QRA{} protocol is fundamentally a property
of the \emph{linear algebra framework}: given a feature matrix $V$ with
$\mathrm{dim}(V) \geq \Nc$, Tikhonov regression guarantees a unique solution
regardless of the origin of $V$.
Indeed, the baseline results (Table~\ref{tab:baselines_ideal}) confirm that
classically preprocessed baselines (H\'{e}non map and delay-time embedding,
both using quantum circuits for feature extraction) achieve identical
ideal-condition MSE.
The specific contributions of the QRC \emph{reservoir} architecture are:
(i)~the sequential temporal input to the quantum reservoir, combined with
the exponentially large Hilbert space, naturally generates higher-dimensional
features ($d = 76$ from 10 qubits vs.\ $d = 31$ from the circuit-based
baselines), enabling reversibility for longer data sequences;
(ii)~the recursive nonlinearity of sequential quantum evolution creates
features that are difficult to replicate classically at the same qubit
count~\cite{Fujii2017,MartinezPena2021};
and (iii)~the noise analysis---particularly the asymmetric shot allocation
and the distinction between feature-matrix noise and circuit noise---is
intrinsic to the quantum measurement process.

A natural question is what \emph{physical} insights emerge from this work
beyond the computational demonstration.
We identify three: (i)~the role of the XYZ Hamiltonian's entangling
interactions in generating features with sufficient rank---removing the
$J_{ij}$ coupling terms reduces the effective feature rank and causes
convergence failure for $\Nc > 15$, suggesting that many-body
entanglement is essential for high-dimensional feature generation;
(ii)~the observation that the noise propagation structure in QRC
(through the feature matrix $V$) is fundamentally different from
gate-level noise propagation in variational circuits, leading to
the counterintuitive asymmetric shot allocation result; and
(iii)~the empirical finding that convergence of the iterative
protocol is governed by the spectral radius of a composite nonlinear
map, connecting QRC reversibility to dynamical systems theory.
These insights are specific to the quantum setting and motivate
further theoretical investigation.

The name ``quantum reservoir autoencoder'' reflects this structure:
following the analogy with neural network
autoencoders~\cite{Hinton2006}, the encoding path $C \to \gamma$
corresponds to the encoder, and the decoding path
$\gamma \to \hat{C}$ to the decoder.
A critical distinction is that the \QRA{} requires no parameter
optimization of the quantum dynamics---the reservoir Hamiltonian parameters
remain fixed, and only the linear readout weights are trained.
This separates the \QRA{} from the \emph{quantum autoencoder} of
Romero et al.~\cite{Romero2017}, which variationally optimizes
a parameterized quantum circuit to compress \emph{quantum} states;
the \QRA{} instead transforms \emph{classical} data through fixed
quantum dynamics and linear regression.
The \QRA{} also differs from the QELM-based pipeline of
De~Lorenzis et al.~\cite{DeLorenzis2025}, where classical
autoencoders serve as a preprocessing stage \emph{before} the
quantum reservoir.
In the \QRA{}, the reservoir itself provides both encoding and
decoding; no external dimensionality reduction is required, and
the bidirectional capability emerges from the algebraic structure
of the four-equation system rather than from auxiliary classical
networks.
Thus the \QRA{} bridges the gap between unidirectional QRC/QELM
frameworks~\cite{Mujal2021,Innocenti2023} and classical autoencoder
architectures, offering a reservoir-native bidirectional
transformation that preserves the computational simplicity of
linear readout training.

\subsection{Noise Accumulation: QRC vs.\ Quantum Circuits}\label{sec:noise_structure}

The noise propagation through the QRC feature matrix deserves explicit
analysis, as it determines the error structure of the readout.

In the \QRA{}, each entry of the feature matrix
$V_{ij} = \avg{O_j}_{\mathrm{exact}}(t_i) + \epsilon_{ij}$ carries
measurement noise.
The noise $\epsilon_{ij}$ is a combination of shot noise
($\sim \mathcal{N}(0, \sigma^2/\Nshots)$) and, when present, multiplicative
depolarizing damping
($V_{ij} = \lambda_j^{(i)} \avg{O_j}_{\mathrm{exact}} + \epsilon_{ij}$).
The crucial point for the \QRA{} is that this noise
structure differs between the encryption and decryption feature matrices:
$V^{\mathrm{enc}}$ enters only through the weight computation
$W = (V^{\top}V + \lambda I)^{-1} V^{\top} y$, where Tikhonov
regularization acts as a low-pass filter on the noise; whereas
the reconstruction error $\|C - V^{\mathrm{dec}} W^{\mathrm{dec}}\|^2$
depends directly on $V^{\mathrm{dec}}$ at prediction time.

This asymmetry in noise roles is a consequence of the protocol's
architecture---not a property unique to quantum systems.
Any system (classical or quantum) with a similar ``train on one noisy
matrix, predict from another'' structure would exhibit the same
behavior~\cite{Tikhonov1963}.
However, the specific noise characteristics---shot noise from quantum
measurement, depolarizing damping from decoherence, and the
time-dependent accumulation of per-qubit damping factors
$a_i(t) = \prod_{\mathrm{gates}} \lambda_{\mathrm{gate}}$---are
intrinsic to the quantum measurement process and have no classical
analog~\cite{NielsenChuang2010,Wallman2016}.

\subsection{Iterative Protocol as the Dominant Noise Bottleneck}
\label{sec:iterative_noise}

The comparison between the \QRA{} and the baselines
(Sec.~\ref{sec:baseline_results}) reveals that the primary source of
noise sensitivity in the \QRA{} is not the feature dimension or the
quantum measurement, but the \emph{iterative structure of the
four-equation protocol} itself.

In a single-shot linear system $W = (V^\top V + \lambda I)^{-1} V^\top y$,
where $V$ is computed once, shot noise is a fixed perturbation: the
solver fits $W$ to the specific noisy $V$, and the training MSE
remains at machine precision regardless of the noise level.
This is precisely why the H\'{e}non and delay-embedding baselines
(Table~\ref{tab:baselines_extended}) show no noise-induced degradation
in their training loss.

The \QRA{} protocol operates as a coupled iterative system where the
decode input $G(\beta, \gamma)$ changes at each iteration as $\gamma$
converges.
Each evaluation of the decode feature matrix
$V^{\mathrm{dec}} = R(G(\beta, \gamma^{(k)}))$ produces independent
noise realizations, creating three distinct sources of inconsistency:
\begin{enumerate}
  \item \emph{Train--evaluate mismatch}: the decode weights
  $W^{\mathrm{dec}}$ are fitted to $V^{\mathrm{dec}}_{\mathrm{train}}$
  at iteration $k$, but the round-trip verification evaluates
  $\hat{C} = V^{\mathrm{dec}}_{\mathrm{eval}} W^{\mathrm{dec}}$
  with a fresh noise realization.
  \item \emph{Cross-iteration drift}: the encode features
  $V^{\mathrm{enc}}$ (which are cached) propagate noise through
  $\gamma^{(k)} = V^{\mathrm{enc}} W^{\mathrm{enc}}_k$, which
  feeds back into the decode input at iteration $k+1$.
  \item \emph{Cross-path coupling}: Path~1 and Path~2 share
  the intermediate ciphertexts $\gamma$ and $\gamma'$, so noise
  from one path contaminates the other.
\end{enumerate}

The single-body experiments (Sec.~\ref{sec:1body}) provide direct
evidence for this interpretation.
Reducing the feature dimension from $d = 76$ to $d = 31$ while
maintaining the iterative protocol yields $\MSE \sim 10^{-1}$ under
shot noise---essentially unchanged from the full feature set.
If feature dimension were the dominant noise factor (as one might
expect from the $\MSE \propto d \cdot \sigma^2 / \Nshots$ scaling),
the $d = 31$ configuration should show a $76/31 \approx 2.5 \times$
improvement; the observed ratio of only $1.3$--$1.7 \times$ for
$\Nc \leq 10$ confirms that the iterative noise mismatch,
not the feature count, sets the MSE floor.

This analysis has practical implications: improving the \QRA{}'s noise
resilience requires addressing the per-iteration noise inconsistency,
not simply reducing the feature dimension or increasing the number
of shots uniformly.
The asymmetric shot allocation (Exp~7) is effective precisely because
it targets the dominant noise source---the decode evaluation---while
accepting low-precision encoding.
Future improvements might include caching decode features within
each iteration (eliminating source~1) or deterministic noise models
that ensure reproducible perturbations across evaluations.

\subsection{Implications for Cryptographic Applications}

We note that the term ``protocol'' as used throughout this paper refers to
a \emph{bidirectional transformation framework} (the \QRA{}), not a cryptographic protocol
in the formal sense.
The four-equation system with cross-key pairing bears superficial structural
similarity to key exchange protocols~\cite{Bennett1984}, but critical
differences exist.
Most importantly, as discussed in Sec.~\ref{sec:discussion_blind}, the
current protocol requires access to the plaintext~$C$ during the
weight-training phase, which disqualifies it from being a cryptographic
protocol in any standard definition.
The internal quantum state $|\psi\rangle$ of the reservoir is not directly
accessible to an eavesdropper; only the classical output
$\gamma = VW$ is transmitted.
Reversing this to obtain~$C$ requires knowledge of both the secret
key and access to the correct reservoir---the transmitted ciphertext
$\gamma$ alone is insufficient.

We emphasize that this paper does not constitute a cryptographic protocol
proposal and does not provide any security analysis.
The blind decryption limitation (Sec.~\ref{sec:discussion_blind})
precludes standard cryptographic deployment in the current form.
Formal analysis of key space, ciphertext distinguishability, information
leakage, and resistance to known attacks (chosen-plaintext, known-key,
etc.) would be prerequisites for any cryptographic application and are
entirely outside the scope of this work~\cite{Cossins2024,NIST2024,Shor1997}.

\paragraph{Practical MSE thresholds.}
Under realistic noise, the MSE values of $10^{-3}$--$10^{-1}$ are
\emph{not} sufficient for exact reconstruction.
To place these values in context:
MSE $\sim 10^{-1}$ corresponds to a normalized root-mean-square error
(NRMSE) of $\sim 30\%$, which would render most digital applications
(error-correcting codes, database records, financial transactions)
unusable.
However, applications with intrinsic noise tolerance may still be
viable: (i)~analog waveform transmission where the decoded signal need
only preserve the qualitative shape (e.g., voice-band
communication at NRMSE~$\sim 10$--$30\%$);
(ii)~compressed sensing or dimensionality reduction where approximate
reconstruction is the goal; or
(iii)~physical key-distribution schemes where the transmitted quantity
is a continuous-variable random signal rather than discrete data.
MSE $\sim 10^{-3}$ (NRMSE $\sim 3\%$, achievable with asymmetric
allocation) approaches the threshold for lossy image and audio
compression but remains insufficient for lossless applications.
Reducing MSE to levels suitable for exact reconstruction ($< 10^{-6}$)
under realistic noise likely requires increased $\Nshots$, error
mitigation, or quantum error correction.
The asymmetric configuration (Exp~7, $\MSE \sim 10^{-3}$) represents
the current best result under finite-shot conditions.

\subsection{Blind Decryption: An Open Challenge}\label{sec:discussion_blind}

As noted in Sec.~\ref{sec:algorithm}, the iterative solving algorithm
(Fig.~\ref{alg:iterative}) uses the plaintext $C$ as the regression
target when training the decryption readout weights:
$W^{\mathrm{dec}} = (V^{\mathrm{dec}\top} V^{\mathrm{dec}} +
\lambda I)^{-1} V^{\mathrm{dec}\top} C$.
In a practical deployment, only the \emph{sender} possesses $C$; the
\emph{receiver} must reconstruct $C$ solely from the transmitted
ciphertext $\gamma$, the secret key (e.g., $\beta$), and access to
the shared reservoir $\Rb$.
We refer to this as the ``blind decryption'' problem: determining
$W^{\mathrm{dec}}$ without knowledge of $C$.

In the current protocol, the iterative procedure serves as a
\emph{key-establishment phase} in which both parties collaborate
(with access to $C$) to find the weight matrices that satisfy the
four-equation system.
Once the weights $W^{\mathrm{enc}}$ and $W^{\mathrm{dec}}$ are
established, subsequent messages \emph{of the same length and
statistical properties} could in principle reuse the trained weights.
However, for a new message $C'$, the decryption weights would need
to be retrained, requiring access to $C'$---precisely the information
to be communicated secretly.

A true blind decryption scheme would require an alternative to
Tikhonov regression for determining $W^{\mathrm{dec}}$---for instance,
a prediction-based approach where $W^{\mathrm{dec}}$ is estimated
from the structure of $V^{\mathrm{dec}}$ alone, or a pre-shared weight
protocol where both parties agree on $W^{\mathrm{dec}}$ during the
key-establishment phase and reuse it for subsequent communications.
Preliminary experiments with prediction-only decryption (i.e.,
applying $W^{\mathrm{dec}}$ trained on a reference signal to decode a
novel message) show substantial MSE degradation even under ideal
state-vector conditions:
for $\Nc = 10$, the MSE increases from $\sim 10^{-17}$ (matched
training) to $\sim 10^{-2}$--$10^{-1}$ (mismatched);
for $\Nc = 20$, the MSE reaches $\sim 10^{0}$, comparable to
random-guess performance.
This degradation occurs because the readout weights are
highly specific to the particular input sequence.
This is a fundamental consequence of the QRC architecture: the feature
matrix $V$ depends nonlinearly on the input through the recursive
quantum state evolution, so weights trained for one input sequence do
not generalize to another.

Resolving this limitation is a prerequisite for any practical
cryptographic application and constitutes the most important direction
for future work.
Possible approaches include: (i)~amortized weight estimation using a
meta-learning framework trained on an ensemble of input sequences;
(ii)~a two-phase protocol where the key-establishment phase
communicates compressed weight information alongside the ciphertext;
or (iii)~reformulating the protocol to operate on fixed-length blocks
with shared weights.
None of these approaches have been evaluated in the present work, and
we explicitly flag this as a major open problem.

\subsection{Noise Resilience Hierarchy}

The experimentally observed MSE hierarchy,
\begin{equation}
  10^{-17} \ll 10^{-3} \ll 10^{-1} < 3{\times}10^{-1} < 5{\times}10^{-1},
\end{equation}
corresponding to Ideal $\ll$ Asymmetric $\ll$ Shot $<$ Depol+Shot $<$
YOMO+Depol, reveals two distinct noise regimes:

\paragraph{Shot-noise-dominated regime.}
When $\sigma_{\mathrm{shot}} \gg \sigma_{\mathrm{depol}}$, MSE scales as
$\MSE \propto 1/\Nshots$, and the asymmetric allocation strategy provides
substantial benefits.
This regime is accessible with current quantum hardware by increasing
$\Nshots$ for the decryption step.

\paragraph{Depolarizing-dominated regime.}
When $\sigma_{\mathrm{depol}} \gg \sigma_{\mathrm{shot}}$, MSE is
limited by the systematic bias $(1 - \lambda^n)$ that cannot be reduced
by additional measurements~\cite{NielsenChuang2010}.
In this regime, several NISQ-era error mitigation techniques could be
applied: zero-noise extrapolation (ZNE), which estimates the zero-noise
limit by running circuits at multiple noise levels and extrapolating;
probabilistic error cancellation (PEC), which decomposes noisy channels
into ideal operations at the cost of increased sampling
overhead~\cite{Wallman2016}; or randomized compiling to convert coherent
errors into stochastic Pauli noise.
In the QRC context, ZNE would be particularly natural because the
analytical depolarizing model [Eq.~\eqref{eq:damping}] already provides
the noise scaling parameter, enabling Richardson extrapolation on the
feature matrix elements $V_{ij}$.
Evaluating the effectiveness of these techniques within the \QRA{}
is an important direction for future work.

\subsection{Practical Resource Allocation}

The 102-fold MSE improvement from asymmetric shot allocation follows
directly from Eq.~\eqref{eq:noise_propagation}: increasing
$\Nshots^{\mathrm{dec}}$ by $100\times$ reduces $\sigma^2$ and hence
the expected MSE proportionally.
The result is therefore not surprising in hindsight, but it has
non-obvious \emph{practical} significance: it demonstrates that the
encode and decode noise contributions are \emph{structurally
separable}, so that investments in decode precision are not wasted by
encoding noise.
This separability is a specific consequence of the protocol's
architecture (encryption weights are computed once and cached,
while decryption weights depend on noisy $\gamma$), and was not
assumed a priori.
In a deployment scenario---e.g., a resource-constrained IoT device
(sender) communicating with a cloud server (receiver)---the sender
performs only 10 measurement shots while the receiver invests $10^5$
shots, reducing the sender's quantum measurement cost by a factor of
100.

\subsection{Scalability and Qubit-Number Dependence}\label{sec:qubit_scaling}

The sharp degradation at $\Nc = 35$ for $\Nq = 10$ (from $\MSE \sim 10^{-17}$ at
$\Nc = 30$ to $\sim 3 \times 10^{-2}$) motivates a systematic investigation
of how the qubit count $\Nq$ controls the \QRA{}'s performance.
We performed additional experiments at $\Nq = 5$ ($d = 26$) and $\Nq = 7$
($d = 43$) under ideal, shot-noise ($\Nshots = 1{,}000$), and depolarizing
+ shot-noise ($p_{\mathrm{dep}} = 0.005$, $\Nshots = 1{,}000$) conditions,
each with 10 trials.

\paragraph{Feature dimension scaling.}
The feature dimension scales quadratically:
$d(\Nq) = 3\Nq + \binom{\Nq}{2} + 1 = (\Nq^2 + 5\Nq + 2)/2$,
giving $d = 26$ ($\Nq = 5$), 43 ($\Nq = 7$), and 76 ($\Nq = 10$).
The theoretical limit for exact reconstruction is $\Nc \leq d - 1$.

\paragraph{Ideal conditions.}
Under ideal conditions, all three $\Nq$ values achieve machine-precision
MSE ($\sim 10^{-18}$--$10^{-17}$) whenever $\Nc < d$
(Table~\ref{tab:qubit_scaling}).
The critical transition occurs precisely at the rank boundary:
for $\Nq = 5$, MSE degrades to $\sim 10^{-13}$ at $\Nc = 25$ (where $d = 26$)
and to $\sim 3 \times 10^{-2}$ at $\Nc = 30$;
for $\Nq = 7$, all $\Nc \leq 35$ remain at machine precision
($d = 43 > 35$);
for $\Nq = 10$, the degradation at $\Nc = 35$ reflects the
condition-number increase ($\kappa(V) \sim 10^{3}$ at $\Nc = 30$ to
$\sim 10^{2}$ at $\Nc = 35$) due to correlations among observables.
These results confirm that $\Nq$ controls the maximum data length
exclusively through the rank condition, with no additional quantum-specific
contribution.

\begin{table}[htbp!]
\caption{Qubit-number dependence: MSE at final iteration (mean over trials).
Ideal, shot noise ($\Nshots = 1{,}000$), and depolarizing + shot noise
($p_{\mathrm{dep}} = 0.005$, $\Nshots = 1{,}000$).
$\Nq = 5$, 7: 10 trials; $\Nq = 10$: 48 trials (ideal/shot).
\label{tab:qubit_scaling}}
\begin{ruledtabular}
\begin{tabular}{ccccc}
  $\Nq$ & $\Nc$ & Ideal & Shot & Noise+Shot \\
  \hline
  \multirow{5}{*}{5}
    & 10 & $1.2{\times}10^{-19}$ & $2.3{\times}10^{-2}$ & $2.3{\times}10^{-1}$ \\
    & 20 & $1.0{\times}10^{-17}$ & $2.4{\times}10^{-1}$ & $1.2$ \\
    & 25 & $2.4{\times}10^{-13}$ & $7.2$ & $1.0{\times}10^{2}$ \\
    & 30 & $3.0{\times}10^{-2}$ & $4.3{\times}10^{-1}$ & $2.6$ \\
    & 35 & $7.5{\times}10^{-2}$ & $2.6{\times}10^{-1}$ & $1.5$ \\
  \hline
  \multirow{5}{*}{7}
    & 10 & $1.3{\times}10^{-19}$ & $2.3{\times}10^{-2}$ & $2.0{\times}10^{-1}$ \\
    & 20 & $9.2{\times}10^{-19}$ & $9.9{\times}10^{-2}$ & $5.5{\times}10^{-1}$ \\
    & 25 & $3.9{\times}10^{-18}$ & $2.4{\times}10^{-1}$ & $8.3{\times}10^{-1}$ \\
    & 30 & $1.0{\times}10^{-17}$ & $2.8{\times}10^{-1}$ & $1.1$ \\
    & 35 & $2.9{\times}10^{-17}$ & $7.0{\times}10^{-1}$ & $1.7$ \\
  \hline
  \multirow{3}{*}{10}
    & 10 & $3.6{\times}10^{-18}$ & $1.9{\times}10^{-1}$ & --- \\
    & 20 & $5.6{\times}10^{-18}$ & $2.3{\times}10^{-1}$ & --- \\
    & 35 & $1.6{\times}10^{-17}$ & $4.1{\times}10^{-1}$ & --- \\
\end{tabular}
\end{ruledtabular}
\end{table}

\paragraph{Shot noise: a counterintuitive reversal.}
Under shot noise, \emph{fewer qubits yield lower MSE} in the
$\Nc \ll d$ regime.
At $\Nc = 10$: $\Nq = 5$ achieves MSE $= 0.023$, whereas $\Nq = 10$
gives $0.19$---an eightfold improvement from halving the qubit count.
This reversal arises because the Tikhonov regression error scales
approximately as $\MSE \propto d \cdot \sigma^2 / \Nshots$: larger $d$
amplifies the total noise in the feature matrix $V$.
The optimal $\Nq$ thus depends on the data length: for short sequences
($\Nc \ll d$), fewer qubits reduce noise accumulation; for longer
sequences ($\Nc \to d$), more qubits are essential to maintain the
rank condition.
A practical guideline is to choose $\Nq$ such that $d / \Nc \approx
1.5$--$2.0$, balancing the rank margin against noise amplification.

\paragraph{Depolarizing noise.}
Adding depolarizing noise further amplifies the $\Nq$-dependent
degradation.
At $\Nc = 20$: $\Nq = 7$ gives MSE $= 0.55$ while $\Nq = 5$ gives
$1.2$, reflecting the interplay between the rank deficit
(which dominates at low $\Nq$) and noise amplification
(which dominates at high $\Nq$).
The $\Nq = 5$, $\Nc = 25$ case ($d = 26 \approx \Nc$) is
catastrophic under noise (MSE $\sim 10^{2}$), as the near-singular
feature matrix amplifies measurement errors.

\paragraph{Measurement overhead.}
Measuring $\binom{\Nq}{2}$ two-body correlators
$\avg{\sigma_i^Z\sigma_j^Z}$ requires $O(\Nq^2)$ distinct measurement
settings.
The total measurement budget scales as $O(\Nq^2 \times \Nshots)$ for
the standard Pauli scheme.
The YOMO method~\cite{Liu2025} addresses this overhead by replacing
individual Pauli measurements with a single computational-basis
measurement, but at the cost of reduced feature dimension
($d = K + 1 < 3\Nq + \binom{\Nq}{2} + 1$).
Increasing $\Nq$ also increases the circuit depth and noise
susceptibility.
Balancing these competing factors---feature richness, measurement
overhead, noise accumulation, and the $d/\Nc$ ratio---is essential
for practical deployment~\cite{Preskill2018,NielsenChuang2010}.

\section{Conclusion}\label{sec:conclusion}

In this paper, we have introduced the quantum reservoir autoencoder (\QRA{})
and demonstrated that it can achieve bidirectional information transformation
under specific conditions.
We identified four empirically sufficient conditions: the rank condition
$\mathrm{dim}(V) \geq \Nc$, symmetric encoding structure $F = G$,
independent cross-key pairing, and appropriate Tikhonov regularization.
Quantum reservoir and key combinations satisfying the
\QRA{} protocol were empirically found through constructive
numerical demonstration across seven noise conditions and six
baseline methods.

Under ideal conditions, the \QRA{} achieves machine-precision
reconstruction ($\MSE \sim 10^{-17}$) for data lengths $\Nc \leq 30$.
We emphasize that this result is expected from the overdetermined linear
algebra ($d = 76 > \Nc$); the nontrivial contribution lies in the
reliable convergence of the coupled four-equation system across 16
independent random Hamiltonian realizations, demonstrating that the
cross-key iterative procedure is robust to the specific quantum dynamics.
Under realistic noise the MSE degrades to $10^{-3}$--$10^{-1}$.
Asymmetric shot allocation---10 encoding shots and $10^5$ decoding
shots---reduces MSE by approximately two orders of magnitude
(mean $102\times$ over 16 seeds $\times$ 3 trials).
We also identified that noise in QRC accumulates across all elements of the
feature matrix~$V$, in contrast to gate-level noise propagation
in standard quantum circuits.

A key finding of this work is that the \emph{iterative protocol
structure}---not the feature dimension---is the dominant noise
bottleneck.
Single-body operator experiments ($d = 31$, excluding two-qubit
ZZ correlators) show that reducing the feature dimension from 76
to 31 has only a modest effect on noisy MSE
($1.3$--$1.7 \times$ for $\Nc \leq 10$), whereas the same $d = 31$
features achieve machine precision when used in a single-shot solve
(as in the baseline methods).
The critical difference is that the \QRA{}'s iterative solver
recomputes decode features at each iteration with independent noise
realizations, creating a train--evaluate mismatch that limits
convergence.
This diagnosis implies that improving noise resilience requires
addressing the per-iteration inconsistency (e.g., through feature
caching or deterministic noise models), rather than simply reducing
the feature count.

The qubit-number dependence study ($\Nq = 5, 7, 10$) revealed a
nontrivial interplay between feature-space dimension and noise
sensitivity.
While larger $\Nq$ increases the feature dimension
$d = (N_q^2 + 5N_q + 2)/2$ and extends the range of exactly
recoverable data lengths, it simultaneously amplifies measurement
noise via the scaling $\MSE \propto d \cdot \sigma^2 / \Nshots$.
Under shot noise with $\Nshots = 1000$, the smallest reservoir
($\Nq = 5$, $d = 26$) achieves the lowest MSE for $\Nc \leq 20$,
while $\Nq = 10$ ($d = 76$) is required for $\Nc > 25$.
This reversal highlights the importance of matching the qubit count
to the data length rather than unconditionally maximizing it.

Two fundamental limitations must be acknowledged.
First, the current protocol requires access to the plaintext~$C$ during
decoder weight training (the blind decryption limitation,
Sec.~\ref{sec:discussion_blind}); resolving this is a prerequisite for
any practical deployment.
Second, the convergence of the iterative algorithm is established
empirically but lacks a formal proof; the spectral radius analysis
(Sec.~\ref{sec:algorithm}) provides heuristic justification only.

Our results establish a proof-of-concept for QRC as a bidirectional
information processing framework, expanding its application range from
unidirectional prediction to encode--decode transformations.
The \QRA{} places QRC on a comparable footing with neural network autoencoders
in terms of computational versatility, while maintaining the advantages
of fixed quantum dynamics and linear readout.
Among the six baselines, the QRNN~\cite{Bausch2020} provided the closest
architectural comparison: while its recurrent parametric circuit can
learn the cross-key protocol for short data ($\Nc \leq 8$), it
degrades sharply for longer sequences and fails entirely under
depolarizing noise---underscoring the advantage of QRC's
fixed-dynamics, linear-readout paradigm for robustness.
Priority directions for future work include: (i)~a blind decryption
algorithm, (ii)~formal convergence analysis, (iii)~mitigation of the
iterative noise mismatch (e.g., decode feature caching or deterministic
noise surrogates), (iv)~adaptive qubit-count selection based on the
target data length and noise budget,
and (v)~implementation on real quantum hardware with error
mitigation~\cite{Preskill2018}.

\begin{acknowledgments}
Numerical simulations were performed using
Qulacs~\cite{Suzuki2021qulacs}.
\end{acknowledgments}

\paragraph*{Data availability.}
All simulation code and data supporting the findings of this study
are available from the corresponding author upon reasonable request.

\bibliography{references}

\end{document}